\documentstyle[12pt,amssymbols]{article}  
\input epsf
\newcommand{\bq}{\begin{equation}}  
\newcommand{\eq}{\end{equation}}  
\newcommand{\bqa}{\begin{eqnarray}}  
\newcommand{\eqa}{\end{eqnarray}}  
\newcommand{\ra}{\rightarrow}

\def\half{{1 \over 2}}  
\def\s{\sigma}

\def\D{\Delta}

\def\d{\delta}

\def\ep{\epsilon}

\def\ov{\over}

\def\ed{\end{document}}
\def\ws{\;\;}
\def\db{\bar{d}}
\def\ra{\rightarrow}  
  
\def\2pi{1\over 2\pi i}  

\def\newline{\hfil\break}

\def\ra{\rightarrow}

\def\sq2{\sqrt{2}}  
\def\sqk2{\sqrt{2(k+2}}  
\def\sqk{\sqrt{k}}

\def\be{\begin{equation}}  
\def\ee{\end{equation}}  
\def\br{\begin{array}}  
\def\er{\end{array}}  
\def\bea{\begin{eqnarray}}  
\def\eea{\end{eqnarray}}  
\def\ba{\begin{equation}\begin{array}}
\def\ea{\end{array}\end{equation}}
\def\bac{\begin{equation}\begin{array}{rll}}

\newcommand{\uq}{U_q (\widehat{sl(2)})}

\def\us{\underline{\sigma}}

\def\Z{{\Bbb Z}}
\def\C{{\Bbb C}}

\def\cH{{\cal{H}}}

\def\pl{\prod\limits} 
\def\sl{\sum\limits} 
\def\il{\int\limits}
\def\ft{\infty}

\def\ep{\epsilon}

\begin{document}  
\begin{titlepage} 
\rightline{CRM-2198} 
\rightline{hep-th/9407062}
\rightline{July 14, 1994} 
\vbox{\vspace{15mm}} 
\vspace{1.0truecm} 
\begin{center} 
{\LARGE \bf The Dynamical Correlation Function of the
XXZ Model
}\\[8mm] 
{\large ROBERT A. WESTON$^{1}$  and  A.H. BOUGOURZI$^{2}$}\\ 
[3mm]{\it Centre de Recherche Math\'ematiques, 
Universit\'e de Montr\'eal\\ 
C.P. 6128, Succursale Centre-Ville, Montr\'eal (Qu\'ebec) H3C 3J7, Canada.}\\[15mm] 
\end{center} 
\begin{abstract} 
We perform a spectral decomposition of the dynamical
correlation function of the spin $1/2$ XXZ model
into an infinite sum of products of form
factors. Beneath the four-particle threshold in momentum
space the only non-zero contributions to this sum are
the two-particle term and the trivial vacuum
term. We calculate the
two-particle term by 
making use of the integral expressions for form factors
provided recently by the Kyoto school. We evaluate the
necessary integrals by expanding to twelfth order in $q$.
We show plots of $S(w,k)$, for $k=0$ and $\pi$ at 
various values of the anisotropy parameter, and for
fixed anisotropy at various $k$ around $0$ and $\pi$.
\end{abstract} 
\footnotetext[1]{ 
Email: {\tt westonr@ere.umontreal.ca} 
} 
\footnotetext[2]{ 
Email: {\tt bougourz@ere.umontreal.ca} 
} 
\end{titlepage} 
\section{Introduction}
In the last two years, a remarkable series of papers have
appeared in which the consequences of the quantum affine symmetry
of quantum spin chains is explored \cite{Daval92,collin,idzal93,Fodal93,Jimal93b}. In the simplest case, the
spin chain is the antiferromagnetic spin-1/2 XXZ model, and the
associated non-abelian symmetry that of $\uq$ \cite{Daval92,collin}. 
The $\uq$ symmetry manifests itself in different ways. Firstly,
the Hamiltonian of this model commutes with the algebra
directly. Secondly, the model has a vertex operator 
algebra as a dynamical symmetry. These vertex operators are in turn
intertwiners of certain $\uq$ modules. 
However,
we do not wish to review the work of the Kyoto School
here. An excellent
and accessible reference is the recent book by Jimbo and Miwa 
\cite{JiMi94}.
It is sufficient for our purposes to note that this approach
leads to exact expressions for both correlations functions and
form factors of local operators of the $XXZ$ model. These objects
are given as the trace of vertex operators over irreducible
highest weight representations of $\uq$. Using the free field
representation of the algebra these traces may be evaluated
to yield integral expressions.

Perhaps somewhat surprisingly, one-dimensional quantum spin chains
are relevant to certain physical systems (crystals that have a much
stronger interaction in one direction, making them effectively 
one-dimensional - for  reviews of theoretical approaches
and experimental data, see  \cite{Aff89b,AfSo94} and references
therein). A quantity of
key interest in this context is the dynamical correlation
function (defined in the next section). This quantity is related to
neutron scattering data \cite{Deial90}. In this paper we calculate the dynamical
correlation of the spin-$1/2$ XXZ model by performing a spectral
decomposition, i.e., we re-express the correlation function
as an infinite sum over products of form factors. 
Since we are calculating 
the dynamical correlation function in momentum space, all
terms other than the vacuum and two-particles one vanish if we restrict ourselves to
beneath the four-particle threshold (odd particle terms
are always zero). 
This may seem an artificial restriction, but we are lucky in that
experimentalist are, for the most part, only interested in the region around
the two-particle threshold.
The vacuum term is trivial to evaluate. It gives a delta
function peak at $w=0$ and $k=\pi$.

We evaluate the two-particle form factor by expanding the integral
expression of the Kyoto School in $q$ (to twelfth order).
Up to some subtleties about the integration contour (which we explain),
 the coefficients are
then relatively easy to work out.

In Section 2, we define the $XXZ$ model, and describe the
spectral decomposition
of the dynamical correlation function. In Section 3, we write down
the form factor expressions derived by the Kyoto School. We describe
the general method of q-expanding such integrals, and evaluate the
specific integrals required to twelfth order in $q$. 
In Section 4, we present the results of some consistency checks on our
form factors. We calculate all the contributions to the correlation
function, and show graphically how the correlation function
behaves in various regions of  parameter space defined by
energy, momentum and $q$.
Finally, in Section 5, we draw some conclusions.
\section{The Dynamical Correlation Function of the
XXZ Model}
\subsection{The XXZ Hamiltonian}
The Hamiltonian of the spin $1/2$ XXZ Heisenberg quantum 
spin chain is  \cite{tafa79,KiRe87,Aff89,JiMi94} 
\be {\cal H}=-\half \sum_{i=-\infty}^{\infty} (\s_i^1 \s_{i+1}^1 
+\s_i^2 \s_{i+1}^2 + \Delta\s_i^3 \s_{i+1}^3) \label{hamiltonian}, \ee 
where $\D=(q+q^{-1})/2$ is an isotropy parameter. Here $\s_i^j$ are the
Pauli matrices acting at the $i^{\rm th}$ site, with the Hamiltonian acting
formally on the 
infinite tensor product 
\be W= \cdots V \otimes V \otimes V \otimes V \cdots \label{infprod},\eq 
where $V\simeq \C^2$. We consider the model in the massive 
antiferromagnetic phase $\D<-1$, which we  parametrise by 
$-1<q<0$.
\subsection{The dynamical correlation function}
The dynamical correlation function we consider is 
\bac S^i(w,k)&=&\il_{-\infty}^{\infty} { dt }
\sl_{p \in Z} 
e^{i(wt -kp)} S^i(t,p),\\
S^i(t,p)&=& 
{_i{<vac|}} \us (t,p)\cdot \us (0,0) |vac>_i, \label{corr1}\ea 
where $\s^j(t,p)$ refers to $\s^j$ acting on the site at position
$p$ at time $t$ (in Minkowski space), and $i=0,1$ refers to
a choice of one of the two antiferromagnetic boundary conditions.
We may define these boundary conditions by
arbitrarily choosing the eigenvalue of $\s^3$ at the
$j$'th site, 
$s^j=(-1)^{i+j}$, for $|j| >> 1$.
The dynamical correlation function is of physical interest because of its connection
with neutron scattering data \cite{AfSo94,Deial90}. For a physical system described by
the Hamiltonian \ref{hamiltonian}, the cross section for scattering
of neutrons that give up energy $w$ and momentum $k$ is proportional
to $S^i(w,k)$. 
In this paper we describe
a method of calculating this function as an expansion in the parameter
$q$.
\subsection{Decomposition of the identity}
We evaluate \ref{corr1} by performing a spectral decomposition
(a similar method for calculating the dynamical
 correlation function of 
 the $s=1$ XXX model, as approximated by an O(3) non-linear
sigma model,  was used in reference \cite{AfWe92}).
That is, we insert a complete set of energy eigenstates between
the Pauli matrices. The Kyoto School has shown how to construct
these states within the context of the representation theory of
the quantum affine algebra $\uq$. We don't wish to go into excessive
detail about this rather long story, but rather pull results
from the literature as necessary. 
A complete discussion of this subject
is given in reference \cite{JiMi94}.

Jimbo and Miwa have conjectured (and given strong 
plausibility arguments for) the completeness of the following 
decomposition of
the identity in terms of the eigenstates of the Hamiltonian \cite{JiMi94}:
\be {\Bbb I}=\sl_{i=0,1}\sl_{n \ge 0} \sl_{\ep_n,\cdots,\ep_1}
{1 \ov {n !}} \oint \db \xi_1 \cdots \db \xi_n 
 |\xi_n,\cdots,\xi_1>_{{\ep_n,\cdots,\ep_1};i}\;
{_{i;{\ep_1,\cdots,\ep_n}}{<\xi_1,\cdots,\xi_n|}}, \ee
where, $\db \xi= d \xi/(2 \pi i z)$, the contours are around the unit circle, $|\xi_i|=1$, and 
the pseudoparticle `charges' are $\ep_i=\pm.
1$.

The action of the spatial shift operator and
Hamiltonian on multiparticle eigenstates states is given by
\bac T|\xi_1,\cdots,\xi_n>_i &=&\pl_{i=1}^n\tau(\xi_i)^{-1}
|\xi_1,\cdots,\xi_n>_{1-i}, \\
H_{XXZ}|\xi_1,\cdots,\xi_n>_i &=&\sl_{i=1}^n E(\xi_i)|\xi_1,\cdots,\xi_n>_i, \label{states}\ea  
where, 
\bac \tau(\xi)&=& \xi^{-1} {\theta_{q^4} (q \xi^2) \ov \theta_{q^4} (q 
\xi^{-2})}, \\
E(\xi) &=&{ 1-q^2 \ov 2 q} \xi {d \ov d \xi} \log \tau(\xi), \label{enmom}\ea
and $\theta_a(b)=(a;a)_{\ft} (b;a)_{\ft} (a b^{-1};a)_{\ft}$;
$(a;b)_{\ft}=\pl_{n=0}^{\ft} (1-a b^n)$. For later purposes,
we also define $(a;b,c)=\pl_{n,m=0}^{\ft} (1-a b^n c^m)$.
As before, $i=0,1$ corresponds
to a choice of one of the boundary conditions.
The states \ref{states} correspond to the hole states in the
Bethe Ansatz spectrum \cite{Babal83}.

Inserting this expression for the indentity into \ref{corr1},
and making use of the properties,
\bac \s^j (t,p)&=&e^{i\cH t} T^{-p} \s^j (0,0) T^{p} e^{-i\cH t},\\
T|vac>_i &=&|vac>_{1-i}, \ea
 we 
obtain the following expression for the correlation function:
\ba{ll} &S^i(w,k)=\sl_{n \ge 0} S_n^i(w,k),\ws {\rm where}, \\
& S_n^i(w,k)= \sl_p \sl_{\ep_n,\cdots,\ep_1} 
{1 \ov {n !}} \oint \db \xi_1 \cdots \db \xi_n 
\left({e^{ik} \ov \tau(\xi_1) \cdots \tau(\xi_n)}\right)^p
 \d (w-E(\xi_1) \cdots
-E(z_n))\\
&\times {_{i+p}<vac|}\us (0,0)|\xi_n,\cdots,\xi_1>_{\ep_n,\cdots,\ep_1;i+p}\:
{_{i;\ep_1,\cdots,\ep_n}<\xi_1,\cdots,\xi_n|}
\us(0,0)|vac>_i, \ea
where the boundary conditions are understood as modulo 2.
Using the delta function $\D(z)\equiv \sl_{p\in \Z} z^p$, we may rewrite
this expression as
\ba{ll}
& S_n^i(w,k)=\sl_{\ep_n,\cdots,\ep_1} 
{1 \ov {n !}} \oint \db \xi_1 \cdots \db \xi_n 
  \D \left({e^{2 ik} \ov \tau(\xi_1)^2 \cdots \tau(\xi_n)^2}
\right)
 \d (w-E(\xi_1) \cdots
-E(z_n))\\
&\times \left(
{_{i}<vac|}\us (0,0)|\xi_n,\cdots,\xi_1>_{\ep_n,\cdots,\ep_1;i}
\:
{_{i;\ep_1,\cdots,\ep_n}<\xi_1,\cdots,\xi_n|}
\us{(0,0)}|vac>_i \right.\\
& +
\left. \left({e^{ik} \ov  \tau(\xi_1)\cdots \tau(\xi_n)}\right)
{_{1-i}<vac|}\us (0,0)|\xi_n,\cdots,\xi_1>_{\ep_n,\cdots,\ep_1;1-i}\:
{_{i;\ep_1,\cdots,\ep_n}<\xi_1,\cdots,\xi_n|}
\us{(0,0)}|vac>_i \right).
 \label{corr2}\ea
Two of the integrals may be carried out by making use of
the identities,
\bac \oint \left( {d\xi \ov 2\pi i \xi}\right) \d(f(\xi)) g(\xi)
 &=&\sl_{\xi_0}
{ g(\xi_0) \ov |\xi {d \ov d\xi}f(\xi)|_{\xi_0}},\ws \{\xi_0|
f(\xi_0)=0,|\xi_0|=1\},\\
\oint \left( {d\xi \ov 2\pi i \xi}\right) \D(f(\xi)) g(\xi)
 &=&\sl_{\xi_0}
{ g(\xi_0) \ov |\xi {d \ov d\xi} \log f(\xi)|_{\xi_0}},\ws \{\xi_0|
f(\xi_0)=1,|\xi_0|=1\}.
\label{id1}\ea

The first observation we make about \ref{corr2} is that, for a given
choice of $w$ and $k$, the
integral is only non-zero if their exist $\xi_1,\cdots,\xi_n$ such that
both, $\tau(\xi_1)^2 \cdots \tau(\xi_n)^2=e^{2 i k}$, and,
$E(\xi_1)+\cdots+E(\xi_n)=w$ hold (that is, there is momentum 
and energy conservation). 
In particle physics language, $w$ needs to be greater that the
 n-particle threshold. (If we were dealing with free relativistic bosons,
the analogous argument would imply that the energy and momentum conservation conditions could be met only with
$w^2 >k^2 + n^2 m^2$.)
Putting the argument another way, for a given choice of $w$ and $k$,
the series \ref{corr1} terminates.

Charge conservation in this model implies that $S^i_{n\, {\rm odd}}(w,k)$ 
vanishes. (This point is clear within the context of the Kyoto School
approach when form factors are expressed as traces over
vertex operators. Ultimately, the charge conservation
is related to the 
charge conservation in the
Boltzmann weights of the corresponding six vertex model
\cite{bax82}.) 
So up to the four-particle
threshold, the is an exact equality between $S^i(w,k)$ and
$S_0^i(w,k)+S^i_2(w,k)$. The existence and location of the two and four particle thresholds are shown in Figs. 1 and 2. In Fig. 1, we simply choose
a large number of $(\xi_1,\xi_2)$ pairs randomly destributed
over $|\xi_i|=1$, and
plot $w=E(\xi_1)+E(\xi_2)$ versus $k=-i \log(\tau(\xi_1) +
\tau(\xi_2))$ (we evaluate $E(\xi_i)$ and $\tau(\xi_i)$ by
the $q$ expansions given in Section 4). In Fig. 2, we
do the same but with $(\xi_1,\cdots,\xi_4)$. The thresholds
are the lower edges of the two scatter plots.

The vacuum, or zero-particle, term,
is given by
\bac S_0^i(w,k)&=& \D(e^{2 i k}) \d (w)
\left( {_{i}<vac|}\us (0,0)|vac>_i +e^{ik}{_{1-i}<vac|}\us (0,0)|vac>_{1-i} \right) \\ &\cdot &
{_{i}<vac|}\us (0,0)|vac>_i. \label{vac}\ea
Again, charge conservation implies that ${_i{<vac|}} 
\s^{\pm} |vac>_i =0$.
Using a $\Z_2$ symmetry given in \cite{JiMi94}; 
\be {_{i}<vac|}\s^z (0,0)|vac>_i=-\, 
{_{1-i}<vac|}\s^z (0,0)|vac>_{1-i},\ee
we can simplify this expression to give,
\be S_0^i(w,k)= \D(e^{2 i k}) \d (w)
\left({_{i}<vac|}\s^z (0,0)|vac>_i\right)^2(1-e^{ik}). \ee
So the vacuum contribution to $S^i(w,k)$ is a delta function
peak at $k=\pi$. The coefficient ${_{i}<vac|}\s^z (0,0)|vac>_i$ is non-zero, and is in fact the staggered polarization
$=(q^2;q^2)_{\ft}^2/(-q^2;q^2)_{\ft}^2$ derived by Baxter
\cite{Bax73},
and reproduced by the Kyoto School (by explicitly integrating
the relevant integral formula  \cite{collin}). 

In order to calculate the two particle 
contribution, we use the identities \ref{id1} to carry out the two integrals. This gives,
\ba{ll}
& S_2^i(w,k)=\sl_{\ep_2,\ep_1}\sl_{\xi_1,\xi_2} 
{1 \ov {4 c(\xi_1,\xi_2)}} 
\left( {_{i}<vac|}\us|\xi_2,\xi_1>_{\ep_2,\ep_1;i}\:
{_{i;\ep_1,\ep_2}<\xi_1,\xi_2|}
\us|vac>_i \right. \\
& +
\left. 
\left( {e^{ ik} \ov \tau(\xi_1)\tau(\xi_n)}\right)
{_{1-i}<vac|}\us|\xi_2,\xi_1>_{\ep_2,\ep_1;1-i}
{_{i;\ep_1,\ep_2}<\xi_1,\xi_2|}
\us|vac>_i \right).
\label{corr3} \ea
Here, the sum is over all $(\xi_1,\xi_2)$ which are solutions of
$\tau(\xi_1)^2 \tau(\xi_2)^2=e^{2 i k}$ and $E(\xi_1)+E(\xi_2)=w$.
The Jacobian factor $c(\xi_1,\xi_2)$  is  given by
\be c(\xi_1,\xi_2)={1 \ov q-q^{-1}} |\xi_1 E^{\prime}(\xi_1)
E(\xi_2)
-\xi_2 E^{\prime}(\xi_2) E(\xi_1)|. \ee
\section{Evaluation of Form Factors}
In this section we shall give expressions for the
form factors appearing in equation \ref{corr3}.
\subsection{The results of the Kyoto School}
The Kyoto School have shown how to calculate form factors
of the XXZ model as traces of vertex operators 
over infinite-dimensional irreducible highest weight modules of
$\uq$ - see
\cite{JiMi94} and references therein.  By bosonizing the vertex operators,
and representing the highest weight modules in terms of a bosonic
Fock space, they are able to write down explicit expressions 
(generally given in terms of multiple integrals of elliptic
functions) for 
both form factors and correlation functions. 
The following expressions for the form factors of interest are
given in reference \cite{JiMi94}:
\be {_i{<}}vac|\s^{+}|\xi_2,\xi_1>_{-,-;i}=
(-q)^{1-i} \xi_1^{1-i} \xi_2^{2-i} (q^2)_4 (q^4)_4^3
\rho(q)^2 { \gamma(u_2/u_1) \theta_{q^8} (-u_1^{-1} u_2^{-1}
q^{4i})  \ov \theta_{q^4}(-q^3/u_1) \theta_{q^4} (-q^3/u_2) },
\label{ffact1}\ee \ba{ll}
&\hspace*{-9mm}{_i{<}}vac|\s^{z}|\xi_2,\xi_1>_{+,-;i}=
(-q)^{-i} (1-q^{-2}) (q^2)_4^3 (q^4)_4^3 \rho(q)^2
\gamma(u_2/u_1) \\
&\hspace*{15mm}\times \oint\limits_{C} \db v
\oint\limits_{C_++C_-} \db w (qv^{-1}-q^{-1}u_1^{-1}) w^{1-i} (-v)^i 
(-q^{-1} v)_2 (-q^3/v)_2 \\
&\hspace*{15mm}\times { \theta_{q^8}(-q^{4i} v^2/(w^2 u_1 u_2)) \ov (w)_2 (q^2/w)_2 (-q^{-1}v/w)_2 (-q^3 w/v)_2}
\pl_{k=1,2} { \xi_k^{1-i} (-qu_k/w)_4 (-q^3 w/u_k) \ov
(-q u_k)_4(-q^3 u_k^{-1})_4 (u_k/v)_4 (q^{-2} v/u_k)_4 }
,\label{ffact2}\ea
where $u_i =- \xi_i^2$.
We introduce the notation $(a)_n=(a;q^n)_{\ft}$ and
$(a)_{n,m}=(a;q^n,q^m)_{\ft}$. $\gamma$ and
$\rho$ are defined by
\bac \gamma(a)&=&{ (q^4u)_{4,4} (u^{-1})_{4,4} \ov (q^6 u)_{4,4}
(q^2 u^{-1})_{4,4} },\\
\rho(q)&=&{ (q^4)_{4,4} \ov (q^6)_{4,4}}.\ea
The contours $C,C_{\pm}$ are defined by
\ba{lrlll} C:& u_i q^{4 n} &< v < &u_i q^{2- 4n}\ws
&{\rm for}\ws n \ge 0, \\ 
C_{\pm}:&  q^{(2n+1) \pm 1} &<w<& q^{-(2n+1) \pm 1}\ws
&{\rm for}\ws n \ge 0, \ws {\rm and,} \\
& - v q^{2n-1}&<w<& - v q^{-2n-3}\ws
&{\rm for}\ws n \ge 0, \ws{\rm or \ws equiv}.,\\
& - w q^{2n+3}&<v<& - w q^{-2n+1}\ws
&{\rm for}\ws n \ge 0.\ea
We use the $>,<$ signs to imply that the contour
on the left-hand side runs outside or
inside a series of poles. The contour $C$ is not a circle, because
 $|u_i q^2| <|u_i|$, but it is nevertheless
well defined - it simply loops inside the pole at $u_i q^2$. 

The remaining non-zero form factors are defined through the $\Z_2$
symmetry of the theory \cite{JiMi94} by
\bac {_i{<}}vac|\s^-|\xi_2,\xi_1>_{++;i} &=& 
{_{1-i}{<vac|}}\s^+|\xi_2,\xi_1>_{--;1-i}, \\
{_i{<}}vac|\s^z|\xi_2,\xi_1>_{-+;i} &=& 
- {_{1-i}{<vac|}}\s^z|\xi_2,\xi_1>_{+-;1-i}. \label{z2}\ea
All non-zero dual form factors are also defined in terms
of \ref{ffact1} and \ref{ffact2} via the symmetry \cite{JiMi94}
\be {_{i;\ep_1,\ep_2}{<}} \xi_1,\xi_2|{\cal {O}}|vac>_i=
{_i{<vac|}}{\cal {O}}|-q \xi_1,-q \xi_2>_{-\ep_1,\ep_2;i}, \ee
where ${\cal O}$ represent any local operator in the theory.
\subsection{Evaluation of the integral}
To obtain a useful expression for the dynamical correlation 
function \ref{corr3}, we need to evaluate the integral \ref{ffact2}.
It is a daunting task to attempt this directly, and we instead
employ the technique of $q$ expanding the integral. In order to
illustrate some of the subtleties of this technique, we apply
it first to a simpler integral $I(k,l)$, defined by
\be I(k,l)= \oint\limits_{C} \db \xi { 1 \ov (q^k \xi)_4 (q^l/\xi)_4}
,\label{simp}\ee
where the $k$ and $l$ are integers, and the contour is chosen as $C: q^{l+4n} < \xi < q^{-k-4n}$ for 
$n \ge 0$. We consider three generic cases.
\subsubsection*{\underline{$k,l > 0$}}
In this case we can simply expand each term in the products in the
denominator as a series, i.e.,
\be I(k,l)= \oint\limits_{C} \db \xi \pl_{n\geq 0} \sl_{m_1 \geq 0}
(\xi q^{k+4n})^{m_1} \sl_{m_2 \geq 0}(\xi^{-1} q^{l+4n})^{m_2} .\ee
We may rewrite this as,
\be I(k,l)=\oint\limits_{C} \db \xi 
\sl_{n \geq 0} a_n(\xi) q^n, \ee
where for any $n$, $a_n(\xi)$ is a finite polynomial in $\xi$.
We calculate the residue at $\xi=0$ by extracting the constant
term in $a_n(\xi)$. As an example we find,
\be I(2,2)= 1+q^4+3q^{12}+ 12q^{16} + 21q^{20} + 38q^{24} + 63q^{28}
 + 106q^{32} 
\cdots . \ee
\subsubsection*{\underline{$k<0,l > 0$}}
Consider the example $I(-1,3)$, where $C: q^3<\xi<q$.
Now, we cannot expand the 
denominator in positive powers of $q$ as $|\xi/q|<0$ in the
$(1-\xi/q)$ term. To be able to do so we must shift the contour outside
of the $\xi=q$ pole, i.e.,
\be I(-1,3)=-\oint\limits_{C^{\prime}} \db \xi { q \ov 
\xi (1-q/\xi)(q^{3} \xi)_4 (q^3/\xi)_4} + {1 \ov {(q^4)_4 (q^2)_4}}, \ee
where $C^{\prime}:q<\xi<q^{-3}$, and the second term comes
from the residue at $\xi=q$. We can then calculate the 
integral over the contour $C^{\prime}$
by the method used in the previous case.
\subsubsection*{\underline{$k>0,l=0,\ws {\rm or}\ws k=0,l>0$}}
In each of these cases we can $q$ expand the integrand without
problem, but we also obtain a pole at $\xi=1$ which we must 
exclude
or include for the two cases respectively. 

We now show how we have applied this $q$ expansion technique to the
integral \ref{ffact2}. The location of poles in the integrand is 
determined by the following product of terms in the denominator:
\be (w)_2 (q^2/w)_2 (-q^{-1}v/w)_2 (-q^3 w/v)_2 (u_1/v)_4(u_2/v)_4
(q^2 v/u_1)_4 (q^{-2} v/u_2)_4. \ee
We fix $w$ at a point in the $C_+$ or $C_-$ band ($C_{\pm}$ are
genuinely bands, unlike $C$) and carry out the 
$v$ integration first. In order to $q$ expand we must shift the $v$ contour
outside of the $v=q^2 u_2$ and $v=-q w$ poles, in order that
it lies within the bands $|u_i|<|v|<|q^{-2} u_i|$ and 
$|q w|<|v|<|q^{-1} w|$ (we call this contour $C^{\prime}$).
The entire integrand may then be expanded. In the $v$ integration
of this expanded integrand we are left (at any order in $q$)
with  poles at $v=u_1,u_2,0$.
In the subsequent $w$ integration we get poles at $w=0$, and 
also at $w=1$ for the $C_-$ contour. It remains to calculate
the $w$ integrals associated with the residues at the $v=q^2 u_2$
and $v=-q w$ poles. At $v=q^2 u_2$ we get the following $w$ poles
in the denominator,
\be (w)_2 (q^2/w)_2 (-q u_2/w)_2 (-q w/u_2)_2. \ee
However the third factor is partially cancelled by a term in
the numerator leaving us with,
\be (w)_2 (q^2/w)_2 (-q^3 u_2/w)_4 (-q w/u_2)_2.\label{cont1} \ee
In determing the location of the $w$ contour with respect to
the $u_2$ dependent poles we encounter a new subtlety.
The origin of the rather curious original $C$ contour was that
it arose from the sum of two expressions with contours $C_1:
|q^4 u_i|<|v|<|q^2 u_i|$ and $C_2:|u_i| <|v|<|q^{-2} u_i|$
 \cite{JiMi94}.
(We don't wish to go into the origin or form of these two
expressions - yet again we refer the reader to reference 
\cite{JiMi94}.)
Compatibility of these contours with $|wq^3|<|v|<|wq|$, requires
$|q^4 u_i|<|wq|<|u_i|$ and $|u_i|<|wq|<|q^{-4} u_i|$ respectively -
which are of course incompatible with each other.
So in order to work out whether
we should choose $|w|<|u_2/q|$ or $|w|>|u_2/q|$ in the integral
associated with \ref{cont1}, we must determine whether the $v=q^2 u_2$
arises from the part of the original integral 
associated with the 
contour $C_1$ or the part with contour $C_2$. In fact it arises
from the $C_1$ part and so we must choose $|wq|<|u_2|$. This is
convenient, because it means that we can carry out the $q$ expansion
without shifting the contour any further.

Now consider the $w$ integral associated with the residue at the
$v=-q w$ pole.
In this case we are left with the following $w$ dependent terms in
the denominator,
\be \left(-u_1/(qw)\right)_4 \left(-u_2/(qw)\right)_4. \ee
Again, other products cancel with those in the denominator. We find
that these poles arise only from terms associated with the $C_2$
contribution to the original integral. This means that we must choose
$|u_i|<|w q|$. This causes a problem because we cannot then expand
the $(1+u_2/(q w))$ and $(1+u_2/(qw))$ factors in positive powers 
of $q$. Hence we must shift our contour to $|qw|<|u_i|$, which 
requires that we evaluate the residues at two extra poles.

A complete list of the combinations of poles which we must evaluate
in order to obtain a complete $q$ expansion for the form factor
\ref{ffact2} is given in Table 1. The sign $(\pm)$ in the table
indicates the sign of the contribution of a particular residue to 
the overall integral.

\vspace*{5mm}
{\centerline{{\bf Table 1} - Location of residues evaluated in  association with the form factor \ref{ffact2}}}
\vspace*{5mm}
\[\begin{array} {|l|lll|}\hline
v & w &&\\ \hline\hline
0 & (+) 0 & (+) 1& \\ \hline
1 & (+) 0 & (+) 1& \\ \hline
u_1 & (+) 0 & (+) 1& \\ \hline
u_2 & (+) 0 & (+) 1& \\ \hline
q^2 u_2 & (-) 0 & (-) 1& \\ \hline
- q w & (-) 0 &(-) -u_1/q&(-)-u_2/q\\ \hline
\end{array} \]
\vspace*{3mm}

The technique for calculating ${_i{<}}vac|\s^z| -q \xi,-q \xi>_{+-;i}$ is
completely analogous (although not identical as different poles
contribute).
\section{Results}
In practice, as a non-trivial check of our technique,
we calculated each of the following eight integral
form factors
separately: ${_i{<}}vac|\s^z| \xi_2,\xi_1>_{+-;i},$ 
${_i{<}}vac|\s^z| \xi_2,\xi_1>_{-+;i},$ \linebreak
${_i{<}}vac|\s^z| -q \xi_1,-q \xi_2>_{+-;i}$ 
and ${_i{<}}vac|\s^z|-q \xi_1,-q \xi_2>_{-+;i}$ (each for $i=0,1$).
We did this to 6th order in $q$, and found that the 
 $\Z_2$
symmetry relations \ref{z2} held. In addition we found that the dual form factors
were precisely the complex conjugates of the form factors, i.e.,
\be {_i{<}}vac|\s^z| -q \xi_1,-q \xi_2>_{-\ep_1,-\ep_2;i}
={_i{<}}vac|\s^z| \xi_2^{-1},\xi^{-1}>_{\ep_2,\ep_1;i}. \ee
This equality is also true for the exact expression \ref{ffact1}, i.e.,
\be {_i{<}}vac|\s^+| -q \xi_1,-q \xi_2>_{-\ep_1,-\ep_2;i}
={_i{<}}vac|\s^-| \xi_2^{-1},\xi^{-1}>_{\ep_2,\ep_1;i}. \ee
Having established these identities (to sixth order in $q$
at least), there are only
two independent sets of form factors (the members of
each set being related by $\Z_2$ symmetry or conjugation).
We then proceeded to calculate one member from each set,
in practice ${_0{<vac|}}\s^z|-q\xi_1,-q\xi_2>_{+-;0}$ and
${_1{<vac|}}\s^z|-q\xi_1,-q\xi_2>_{+-;1},$ 
to twelfth order in $q$. The results are given in Tables 2 and
3 at the end of the paper. 

Using these results, we calculated the two quantities 
required in the $\s^z$ contribution to \ref{corr3} (to order
$q^{12}$),
\bac f_i^{z\,I}(\xi_1,\xi_2)&=&
{_i{<vac|}} \s^z|\xi_2,\xi_1>_{+-;i}
{_i{<vac|}} \s^z|-q \xi_1,-q \xi_2>_{+-;i}\\
&&+{_i{<vac|}} \s^z|\xi_2,\xi_1>_{-+;i}
{_i{<vac|}} \s^z|-q \xi_1,-q \xi_2>_{-+;i},\\
f_i^{z\,II}(\xi_1,\xi_2)&=&
{_{1-i}{<vac|}} \s^z|\xi_2,\xi_1>_{+-;1-i}
{_i{<vac|}} \s^z|-q \xi_1,-q \xi_2>_{+-;i}\\
&&+{_{1-i}{<vac|}} \s^z|\xi_2,\xi_1>_{-+;1-i}
{_i{<vac|}} \s^z|-q \xi_1,-q \xi_2>_{-+;i}.\ea
$f_i^{z\,I}$ and $f_i^{z\,II}$ are both actually independent
of $i$ because of the $\Z_2$ symmetry. This is also
true for the exact $\s^+$ and $\s^-$ terms,
\bac f_i^{\pm\,I}(\xi_1,\xi_2)&=&
{_i{<vac|}} \s^{\pm}|\xi_2,\xi_1>_{\mp \mp;i}
{_i{<vac|}} \s^{\mp}|-q \xi_1,-q \xi_2>_{\pm \pm;i},\\
f_i^{\pm\,II}(\xi_1,\xi_2)&=&
{_{1-i}{<vac|}} \s^{\pm}|\xi_2,\xi_1>_{\mp \mp;1-i}
{_{i}{<vac|}} \s^{\mp}|-q \xi_1,-q \xi_2>_{\pm \pm;i}.\ea
 Thus $S^i_2(w,k)$
is also independent of $i$.
Furthermore we find, from our explicit results that,
\ba{lll} f_i^{z\,I}(\xi_1,\xi_2) 
&=f_i^{z\,I}(\pm \xi_1,\mp \xi_2)
&=f_i^{z\,I}(\xi_2,\xi_1),\\
f_i^{z\,II}(\xi_1,\xi_2) 
&= - f_i^{z\,II}(\pm \xi_1,\mp \xi_2)
&=  f_i^{z\,II}(\xi_2,\xi_1).\label{prop1}
\ea
Again, the same properties hold for the corresponding
$\s^{\pm}$ form factors.

This later fact is important when we look more closely at
formula \ref{corr3}, which written in terms of the above functions,
becomes,
\ba{ll}
& S_2^i(w,k)=\sl_{\xi_1,\xi_2} 
{1 \ov {4 c(\xi_1,\xi_2)}} \left(2\left(f_i^{+\,I}(\xi_1,\xi_2)+
f_i^{-\,I}(\xi_1,\xi_2)\right)+ f_i^{z\,I}(\xi_1,\xi_2) \right)\\
&+ \left( {e^{ik} \ov \tau(\xi_1)\tau(\xi_2)}\right)
\left(2\left(f_i^{+\,II}(\xi_1,\xi_2)+
f_i^{-\,II}(\xi_1,\xi_2)\right)+ f_i^{z\,II}(\xi_1,\xi_2) \right).
\ea
It would be nice to able obtain  $S_2^i(w,k)$ without performing the rather tedious
task of solving the energy and momentum constraints
\bac \tau (\xi_1)^2\tau(\xi_2)^2&=&e^{2ik},\\
E(\xi_1)+E(\xi_2)&=&w\label{momentum},\ea  
to get
all the  $\xi_1$ and $\xi_2$ roots for a given $w$ and $k$.
This can be done, if given one pair of $\xi_1$ and $\xi_2$,
that give a particular $(w,k)$ (via the energy relation and
$\tau(\xi_1)\tau(\xi_1)=e^{-i k}$), one can get all the other
solutions of the more general constraints \ref{momentum}.
From the explicit form of $\tau(\xi)$ and $E(\xi)$ (either
\ref{enmom} or \ref{mom},
we find that the set of such solutions is, $(\xi_1,\xi_2),
(\pm \xi_1,\mp
\xi_2), (-\xi_2,-\xi_1)$, and another four with $\xi_1 \leftrightarrow
 \xi_2$. 
Given, 
the properties \ref{prop1}, and the fact that $c(\xi_1,\xi_2)=
c(\pm \xi_1,\mp \xi_2)=c(\xi_2,\xi_1)$, we can label
 $S_2^i(w,k)$ in terms of
the original $(\xi_1,\xi_2)$ pair, and write,
\ba{ll}
& S_2^i(\xi_1,\xi_2)=
{2 \ov { c(\xi_1,\xi_2)}} \left(2\left(f_i^{+\,I}(\xi_1,\xi_2)+
f_i^{-\,I}(\xi_1,\xi_2)\right)+ f_i^{z\,I}(\xi_1,\xi_2) \right.\\
&+
\left. 2\left(f_i^{+\,II}(\xi_1,\xi_2)+
f_i^{-\,II}(\xi_1,\xi_2)\right)+ f_i^{z\,II}(\xi_1,\xi_2) \right).\\
\label{corr4}\ea

For possible applications to neutron scattering data, it is useful, to
work out $S_2^i(w,k)$ as a function of $w$ for fixed $k$ values. 
We
do this by numerically solving the momentum
 constraint $\tau (\xi_1)\tau(\xi_2)=e^{i k}$ 
for a range of $(\xi_1,\xi_2)$ pairs, and calculating
the associated $w$ value and $S_2^i(w,k)$ for each pair.
In order to do this we 
identify $\tau(\xi_i)=e^{-i p(\theta_i)}$, $\xi_i=
e^{i \theta_i}$, and use suitably truncated
versions of the 
following $q$ expansions of the functions \ref{enmom} (we calculate $c(\xi_1,\xi_2)$ in the
same fashion):
\bac p(\theta)&=& \theta+ 2 \sl_{m>0} {sin(2 \theta m) q^m
\ov m(1+q^m)} \\
E(\theta)&=& { (q-q^{-1}) \ov 2} \left\{
 \sl_{m>0} {4 cos(2 \theta m)q^m \ov m(1+q^m)} \right\}.\label{mom}\ea

In Figs 3 and 4, we plot $S_2(w,k=0)$ and $S_2(w,k=\pi)$ for a 
range of $q$ values. We find, in each case, that there is a rounded
peak above the two-particle threshold. The location of the
threshold moves to smaller $w$ at $q$ decreases towards $-1$.
This is as expected since, from \ref{mom}, their is a mass gap in the theory,
\be E(p=0)= {(q-q^{-1}) \ov 2}\sl_{m > 0} { 4 q^m \ov {m(1+q^m)}}, \ee
which vanishes as $q \ra -1$. The results at $k=0$ and $k=\pi$
are similar in form, although different in scale. The results at
$k=0$ and $k=\pi$ are different since $S_2^i(\pm \xi_1,\mp \xi_2) \ne
S_2^i(\xi_1,\xi_2)$ - as follows from equation \ref{corr4}.
Our expressions for $S_2(w,k=0)$ and $S_2(w,k=\pi)$ are easily convergent 
at all three $q$ values. As $|q|$ increases, the peaks become infinite
spikes at, or very close to, the threshold 
(at least on the scale of the plots shown).

In Figs 5 and 6, we plot $S_2(w,k)$ for a range of values at around
$k=0$ and $k=\pi$, and at fixed $q=-0.2$. The location of the
peaks move, as the location of the two-particle threshold moves
as in Fig. 1. A rather curious fact is that the peak increases in
height as $k$ is increased from zero, and decreases in height
at $k$ is increased from $\pi$.

\section{Conclusions}
One of the criticisms sometimes made of the recent work on
quantum spin chains, is that, despite all of the mathematical
insights gained along the way, in the end one is left
with  formulae
for correlation functions and form-factors which are 
intractable. We hope we have shown that, at the very least,
these integral formulae can be regarded as generating expressions
for $q$ expansions of physically interesting quantities.
We also hope that we have shown how, by making use of the
 expressions for
form factors, one can use this approach to analyse dynamical quantities -
rather that the short distance equal-time correlation functions to
which it has formerly been applied.
Of course, we have also calculated a quantity which, closer to $q=-1$
may be relevant to the anisotropic spin-chains found in nature.
We hope to comment on such
applications in the future.
\section*{Acknowledgements}
We with to thank Prof.\,\,Miwa for his useful comments and
suggestions. RAW thanks CRM for providing him with a post-doctoral
fellowship. AHB acknowledges and thanks the NSERC for his
post-doctoral fellowship. We both wish to thank all members
of CRM for their encouragement and advice.
\newpage
{
}
\newpage
{\bf Table 2} $\;
\hfill {_{0}{<}}  vac|\s^z|-q \xi_1,-q \xi_2>_{+-;0} =
-{_{1}{<}}vac|\s^z|-q \xi_1,-q \xi_2>_{-+;1} $ \linebreak
\hspace*{30mm}\hfill $={_{0}{<}}vac|\s^z|\xi_2^{-1},\xi_1^{-1}>_{+-;0} =
-{_{1}{<}}\xi|\s^z|\xi_2^{-1},\xi_1^{-1}>_{-+;1} $
\[\begin{array}{|l|l|}\hline
q^{\#}& {\rm coefficient} \\ \hline
1&
{{-2\,\xi_1}\ov {\xi_2}} + {{2\,\xi_2}\ov {\xi_1}}
\\ \hline
2&
{{-2}\ov {\xi_1\,\xi_2}} + {{2\,\xi_2}\ov {{{\xi_1}^3}}}
\\ \hline
3&
{{-2}\ov {\xi_1\,{{\xi_2}^3}}} - 
  {{2\,{{\xi_1}^3}}\ov {{{\xi_2}^3}}} + 
  {{4\,\xi_1}\ov {\xi_2}} + {{2\,\xi_2}\ov {{{\xi_1}^5}}} - 
  {{4\,\xi_2}\ov {\xi_1}} + 
  {{2\,{{\xi_2}^3}}\ov {{{\xi_1}^3}}}
\\ \hline
4&
{{-2}\ov {\xi_1\,{{\xi_2}^5}}} - 
  {{2\,\xi_1}\ov {{{\xi_2}^3}}} + {4\ov {\xi_1\,\xi_2}} + 
  {{2\,\xi_2}\ov {{{\xi_1}^7}}} - 
  {{4\,\xi_2}\ov {{{\xi_1}^3}}} - 2\,\xi_1\,\xi_2 + 
  {{2\,{{\xi_2}^3}}\ov {{{\xi_1}^5}}} + 
  {{2\,{{\xi_2}^3}}\ov {\xi_1}}
\\ \hline
5&
{{-2}\ov {\xi_1\,{{\xi_2}^7}}} - 
  {{2\,\xi_1}\ov {{{\xi_2}^5}}} - 
  {{2\,{{\xi_1}^5}}\ov {{{\xi_2}^5}}} + 
  {2\ov {\xi_1\,{{\xi_2}^3}}} + 
  {{8\,{{\xi_1}^3}}\ov {{{\xi_2}^3}}} - 
  {{2\,\xi_1}\ov {\xi_2}} + {{2\,\xi_2}\ov {{{\xi_1}^9}}} - 
  {{2\,\xi_2}\ov {{{\xi_1}^5}}} + 
  {{2\,{{\xi_2}^3}}\ov {{{\xi_1}^7}}} - 
  {{6\,{{\xi_2}^3}}\ov {{{\xi_1}^3}}} + 
  {{2\,{{\xi_2}^5}}\ov {{{\xi_1}^5}}}
\\ \hline
6&
{{-2}\ov {\xi_1\,{{\xi_2}^9}}} - 
  {{2\,\xi_1}\ov {{{\xi_2}^7}}} + 
  {2\ov {\xi_1\,{{\xi_2}^5}}} - 
  {2\ov {{{\xi_1}^3}\,{{\xi_2}^3}}} + 
  {{2\,\xi_1}\ov {{{\xi_2}^3}}} + 
  {2\ov {{{\xi_1}^5}\,\xi_2}}
  - {2\ov {\xi_1\,\xi_2}} - 
  {{2\,{{\xi_1}^3}}\ov {\xi_2}} + 
  {{2\,\xi_2}\ov {{{\xi_1}^{11}}}} - 
  {{2\,\xi_2}\ov {{{\xi_1}^7}}} + 
  {{4\,\xi_2}\ov {{{\xi_1}^3}}}\\&
 + 4\,\xi_1\,\xi_2 + 
  {{2\,{{\xi_2}^3}}\ov {{{\xi_1}^9}}} - 
  {{6\,{{\xi_2}^3}}\ov {{{\xi_1}^5}}} - 
  {{4\,{{\xi_2}^3}}\ov {\xi_1}} + 
  {{2\,{{\xi_2}^5}}\ov {{{\xi_1}^7}}} + 
  {{2\,{{\xi_2}^5}}\ov {{{\xi_1}^3}}}
\\ \hline
7&
{{-2}\ov {\xi_1\,{{\xi_2}^{11}}}} - 
  {{2\,\xi_1}\ov {{{\xi_2}^9}}} + 
  {2\ov {\xi_1\,{{\xi_2}^7}}} - 
  {{2\,{{\xi_1}^7}}\ov {{{\xi_2}^7}}} - 
  {2\ov {{{\xi_1}^3}\,{{\xi_2}^5}}} + 
  {{2\,\xi_1}\ov {{{\xi_2}^5}}} + 
  {{8\,{{\xi_1}^5}}\ov {{{\xi_2}^5}}} + 
  {2\ov {\xi_1\,{{\xi_2}^3}}} - 
  {{8\,{{\xi_1}^3}}\ov {{{\xi_2}^3}}} + 
  {2\ov {{{\xi_1}^7}\,\xi_2}} 
+ {2\ov {{{\xi_1}^3}\,\xi_2}} \\&+ 
  {{2\,\xi_1}\ov {\xi_2}}   +
  {{2\,\xi_2}\ov {{{\xi_1}^{13}}}} - 
  {{2\,\xi_2}\ov {{{\xi_1}^9}}} - 
  {{4\,\xi_2}\ov {{{\xi_1}^5}}} - {{2\,\xi_2}\ov {\xi_1}} - 
  2\,{{\xi_1}^3}\,\xi_2 + 
  {{2\,{{\xi_2}^3}}\ov {{{\xi_1}^{11}}}} - 
  {{4\,{{\xi_2}^3}}\ov {{{\xi_1}^7}}} + 
  {{6\,{{\xi_2}^3}}\ov {{{\xi_1}^3}}} + 
  {{2\,{{\xi_2}^5}}\ov {{{\xi_1}^9}}} - 
  {{6\,{{\xi_2}^5}}\ov {{{\xi_1}^5}}}\\& + 
  {{2\,{{\xi_2}^5}}\ov {\xi_1}} + 
  {{2\,{{\xi_2}^7}}\ov {{{\xi_1}^7}}}
\\ \hline
8&
{{-2}\ov {\xi_1\,{{\xi_2}^{13}}}} - 
  {{2\,\xi_1}\ov {{{\xi_2}^{11}}}} + 
  {2\ov {\xi_1\,{{\xi_2}^9}}} - 
  {2\ov {{{\xi_1}^3}\,{{\xi_2}^7}}} + 
  {{2\,\xi_1}\ov {{{\xi_2}^7}}} + 
  {2\ov {\xi_1\,{{\xi_2}^5}}} + 
  {4\ov {{{\xi_1}^3}\,{{\xi_2}^3}}} + 
  {2\ov {{{\xi_1}^9}\,\xi_2}} - {4\ov {{{\xi_1}^5}\,\xi_2}} + 
  {{2\,{{\xi_1}^3}}\ov {\xi_2}} + 
  {{2\,\xi_2}\ov {{{\xi_1}^{15}}}} \\&- 
  {{2\,\xi_2}\ov {{{\xi_1}^{11}}}} - 
  {{2\,\xi_2}\ov {{{\xi_1}^7}}} - 
  {{6\,\xi_2}\ov {{{\xi_1}^3}}} - 2\,\xi_1\,\xi_2 + 
  {{2\,{{\xi_2}^3}}\ov {{{\xi_1}^{13}}}} - 
  {{4\,{{\xi_2}^3}}\ov {{{\xi_1}^9}}} + 
  {{10\,{{\xi_2}^3}}\ov {{{\xi_1}^5}}} + 
  {{4\,{{\xi_2}^3}}\ov {\xi_1}} + 
  {{2\,{{\xi_2}^5}}\ov {{{\xi_1}^{11}}}} - 
  {{6\,{{\xi_2}^5}}\ov {{{\xi_1}^7}}} - 
  {{6\,{{\xi_2}^5}}\ov {{{\xi_1}^3}}}\\& + 
  {{2\,{{\xi_2}^7}}\ov {{{\xi_1}^9}}} + 
  {{2\,{{\xi_2}^7}}\ov {{{\xi_1}^5}}}
\\ \hline
9&
{{-2}\ov {\xi_1\,{{\xi_2}^{15}}}} - 
  {{2\,\xi_1}\ov {{{\xi_2}^{13}}}} + 
  {2\ov {\xi_1\,{{\xi_2}^{11}}}} - 
  {2\ov {{{\xi_1}^3}\,{{\xi_2}^9}}} + 
  {{2\,\xi_1}\ov {{{\xi_2}^9}}} - 
  {{2\,{{\xi_1}^9}}\ov {{{\xi_2}^9}}} + 
  {2\ov {\xi_1\,{{\xi_2}^7}}} + 
  {{8\,{{\xi_1}^7}}\ov {{{\xi_2}^7}}} + 
  {4\ov {{{\xi_1}^3}\,{{\xi_2}^5}}} + 
  {{2\,\xi_1}\ov {{{\xi_2}^5}}} - 
  {{8\,{{\xi_1}^5}}\ov {{{\xi_2}^5}}}\\& - 
  {2\ov {{{\xi_1}^5}\,{{\xi_2}^3}}} - 
  {2\ov {\xi_1\,{{\xi_2}^3}}} - 
  {{4\,{{\xi_1}^3}}\ov {{{\xi_2}^3}}} + 
  {2\ov {{{\xi_1}^{11}}\,\xi_2}} - 
  {2\ov {{{\xi_1}^7}\,\xi_2}} - {6\ov {{{\xi_1}^3}\,\xi_2}} - 
  {{8\,\xi_1}\ov {\xi_2}} - {{2\,{{\xi_1}^5}}\ov {\xi_2}} + 
  {{2\,\xi_2}\ov {{{\xi_1}^{17}}}} - 
  {{2\,\xi_2}\ov {{{\xi_1}^{13}}}} - 
  {{2\,\xi_2}\ov {{{\xi_1}^9}}} \\&+ 
  {{6\,\xi_2}\ov {{{\xi_1}^5}}} + {{10\,\xi_2}\ov {\xi_1}} + 
  2\,{{\xi_1}^3}\,\xi_2 + 
  {{2\,{{\xi_2}^3}}\ov {{{\xi_1}^{15}}}} - 
  {{4\,{{\xi_2}^3}}\ov {{{\xi_1}^{11}}}} + 
  {{2\,{{\xi_2}^3}}\ov {{{\xi_1}^7}}} + 
  {{2\,{{\xi_2}^3}}\ov {{{\xi_1}^3}}} + 
  {{2\,{{\xi_2}^5}}\ov {{{\xi_1}^{13}}}} - 
  {{4\,{{\xi_2}^5}}\ov {{{\xi_1}^9}}} + 
  {{6\,{{\xi_2}^5}}\ov {{{\xi_1}^5}}} - 
  {{2\,{{\xi_2}^5}}\ov {\xi_1}} \\&+ 
  {{2\,{{\xi_2}^7}}\ov {{{\xi_1}^{11}}}} - 
  {{6\,{{\xi_2}^7}}\ov {{{\xi_1}^7}}} + 
  {{2\,{{\xi_2}^7}}\ov {{{\xi_1}^3}}} + 
  {{2\,{{\xi_2}^9}}\ov {{{\xi_1}^9}}}
\\ \hline
10&
{{-2}\ov {\xi_1\,{{\xi_2}^{17}}}} - 
  {{2\,\xi_1}\ov {{{\xi_2}^{15}}}} + 
  {2\ov {\xi_1\,{{\xi_2}^{13}}}} - 
  {2\ov {{{\xi_1}^3}\,{{\xi_2}^{11}}}} + 
  {{2\,\xi_1}\ov {{{\xi_2}^{11}}}} + 
  {2\ov {\xi_1\,{{\xi_2}^9}}} + 
  {4\ov {{{\xi_1}^3}\,{{\xi_2}^7}}} + 
  {{2\,\xi_1}\ov {{{\xi_2}^7}}} - 
  {2\ov {{{\xi_1}^5}\,{{\xi_2}^5}}} - 
  {2\ov {\xi_1\,{{\xi_2}^5}}} - 
  {{2\,{{\xi_1}^3}}\ov {{{\xi_2}^5}}}\\& - 
  {4\ov {{{\xi_1}^3}\,{{\xi_2}^3}}} + 
  {{2\,\xi_1}\ov {{{\xi_2}^3}}} + 
  {2\ov {{{\xi_1}^{13}}\,\xi_2}} - 
  {2\ov {{{\xi_1}^9}\,\xi_2}} + {2\ov {{{\xi_1}^5}\,\xi_2}} + 
  {2\ov {\xi_1\,\xi_2}} + {{2\,\xi_2}\ov {{{\xi_1}^{19}}}} - 
  {{2\,\xi_2}\ov {{{\xi_1}^{15}}}} - 
  {{2\,\xi_2}\ov {{{\xi_1}^{11}}}} + 
  {{2\,\xi_2}\ov {{{\xi_1}^3}}} - 2\,{{\xi_1}^5}\,\xi_2 \\&+ 
  {{2\,{{\xi_2}^3}}\ov {{{\xi_1}^{17}}}} - 
  {{4\,{{\xi_2}^3}}\ov {{{\xi_1}^{13}}}} + 
  {{4\,{{\xi_2}^3}}\ov {{{\xi_1}^9}}} - 
  {{10\,{{\xi_2}^3}}\ov {{{\xi_1}^5}}} - 
  {{6\,{{\xi_2}^3}}\ov {\xi_1}} + 
  {{2\,{{\xi_2}^5}}\ov {{{\xi_1}^{15}}}} - 
  {{4\,{{\xi_2}^5}}\ov {{{\xi_1}^{11}}}} + 
  {{10\,{{\xi_2}^5}}\ov {{{\xi_1}^7}}} + 
  {{10\,{{\xi_2}^5}}\ov {{{\xi_1}^3}}} + 
  {{2\,{{\xi_2}^7}}\ov {{{\xi_1}^{13}}}} - 
  {{6\,{{\xi_2}^7}}\ov {{{\xi_1}^9}}} \\&- 
  {{6\,{{\xi_2}^7}}\ov {{{\xi_1}^5}}} + 
  {{2\,{{\xi_2}^7}}\ov {\xi_1}} + 
  {{2\,{{\xi_2}^9}}\ov {{{\xi_1}^{11}}}} + 
  {{2\,{{\xi_2}^9}}\ov {{{\xi_1}^7}}}
\\ \hline\end{array}\]
\newpage
{\bf Table 2} continued.
\[\begin{array}{|l|l|}\hline
q^{\#}& {\rm coefficient} \\ \hline
11&
{{-2}\ov {\xi_1\,{{\xi_2}^{19}}}} - 
  {{2\,\xi_1}\ov {{{\xi_2}^{17}}}} + 
  {2\ov {\xi_1\,{{\xi_2}^{15}}}} - 
  {2\ov {{{\xi_1}^3}\,{{\xi_2}^{13}}}} + 
  {{2\,\xi_1}\ov {{{\xi_2}^{13}}}} + 
  {2\ov {\xi_1\,{{\xi_2}^{11}}}} - 
  {{2\,{{\xi_1}^{11}}}\ov {{{\xi_2}^{11}}}} + 
  {4\ov {{{\xi_1}^3}\,{{\xi_2}^9}}} + 
  {{2\,\xi_1}\ov {{{\xi_2}^9}}} + 
  {{8\,{{\xi_1}^9}}\ov {{{\xi_2}^9}}} - 
  {2\ov {{{\xi_1}^5}\,{{\xi_2}^7}}} \\&- 
  {2\ov {\xi_1\,{{\xi_2}^7}}} - 
  {{2\,{{\xi_1}^3}}\ov {{{\xi_2}^7}}} - 
  {{8\,{{\xi_1}^7}}\ov {{{\xi_2}^7}}} - 
  {4\ov {{{\xi_1}^3}\,{{\xi_2}^5}}} - 
  {{2\,\xi_1}\ov {{{\xi_2}^5}}} + 
  {4\ov {{{\xi_1}^5}\,{{\xi_2}^3}}} + 
  {2\ov {\xi_1\,{{\xi_2}^3}}} + 
  {{2\,{{\xi_1}^3}}\ov {{{\xi_2}^3}}} + 
  {2\ov {{{\xi_1}^{15}}\,\xi_2}} - 
  {2\ov {{{\xi_1}^{11}}\,\xi_2}} - 
  {4\ov {{{\xi_1}^7}\,\xi_2}} \\&+ {2\ov {{{\xi_1}^3}\,\xi_2}} + 
  {{8\,\xi_1}\ov {\xi_2}} + {{2\,{{\xi_1}^5}}\ov {\xi_2}} + 
  {{2\,\xi_2}\ov {{{\xi_1}^{21}}}} - 
  {{2\,\xi_2}\ov {{{\xi_1}^{17}}}} - 
  {{2\,\xi_2}\ov {{{\xi_1}^{13}}}} + 
  {{2\,\xi_2}\ov {{{\xi_1}^9}}} - 
  {{4\,\xi_2}\ov {{{\xi_1}^5}}} - {{6\,\xi_2}\ov {\xi_1}} + 
  2\,{{\xi_1}^3}\,\xi_2 + 
  {{2\,{{\xi_2}^3}}\ov {{{\xi_1}^{19}}}} \\&- 
  {{4\,{{\xi_2}^3}}\ov {{{\xi_1}^{15}}}} + 
  {{4\,{{\xi_2}^3}}\ov {{{\xi_1}^{11}}}} + 
  {{4\,{{\xi_2}^3}}\ov {{{\xi_1}^7}}} - 
  {{2\,{{\xi_2}^3}}\ov {{{\xi_1}^3}}} + 2\,\xi_1\,{{\xi_2}^3} + 
  {{2\,{{\xi_2}^5}}\ov {{{\xi_1}^{17}}}} - 
  {{4\,{{\xi_2}^5}}\ov {{{\xi_1}^{13}}}} + 
  {{2\,{{\xi_2}^5}}\ov {{{\xi_1}^9}}} - 
  {{2\,{{\xi_2}^5}}\ov {{{\xi_1}^5}}} - 
  {{4\,{{\xi_2}^5}}\ov {\xi_1}} + 
  {{2\,{{\xi_2}^7}}\ov {{{\xi_1}^{15}}}} \\&- 
  {{4\,{{\xi_2}^7}}\ov {{{\xi_1}^{11}}}} + 
  {{6\,{{\xi_2}^7}}\ov {{{\xi_1}^7}}} - 
  {{4\,{{\xi_2}^7}}\ov {{{\xi_1}^3}}} + 
  {{2\,{{\xi_2}^9}}\ov {{{\xi_1}^{13}}}} - 
  {{6\,{{\xi_2}^9}}\ov {{{\xi_1}^9}}} + 
  {{2\,{{\xi_2}^9}}\ov {{{\xi_1}^5}}} + 
  {{2\,{{\xi_2}^{11}}}\ov {{{\xi_1}^{11}}}}
\\ \hline
12&
{{-2}\ov {\xi_1\,{{\xi_2}^{21}}}} - 
  {{2\,\xi_1}\ov {{{\xi_2}^{19}}}} + 
  {2\ov {\xi_1\,{{\xi_2}^{17}}}} - 
  {2\ov {{{\xi_1}^3}\,{{\xi_2}^{15}}}} + 
  {{2\,\xi_1}\ov {{{\xi_2}^{15}}}} + 
  {2\ov {\xi_1\,{{\xi_2}^{13}}}} + 
  {4\ov {{{\xi_1}^3}\,{{\xi_2}^{11}}}} + 
  {{2\,\xi_1}\ov {{{\xi_2}^{11}}}} - 
  {2\ov {{{\xi_1}^5}\,{{\xi_2}^9}}} - 
  {2\ov {\xi_1\,{{\xi_2}^9}}} - 
  {{2\,{{\xi_1}^3}}\ov {{{\xi_2}^9}}} \\&- 
  {4\ov {{{\xi_1}^3}\,{{\xi_2}^7}}} - 
  {{2\,\xi_1}\ov {{{\xi_2}^7}}} + 
  {4\ov {{{\xi_1}^5}\,{{\xi_2}^5}}} + 
  {2\ov {\xi_1\,{{\xi_2}^5}}} + 
  {{4\,{{\xi_1}^3}}\ov {{{\xi_2}^5}}} - 
  {2\ov {{{\xi_1}^7}\,{{\xi_2}^3}}} + 
  {2\ov {{{\xi_1}^3}\,{{\xi_2}^3}}} - 
  {{2\,\xi_1}\ov {{{\xi_2}^3}}} - 
  {{2\,{{\xi_1}^5}}\ov {{{\xi_2}^3}}} + 
  {2\ov {{{\xi_1}^{17}}\,\xi_2}} - 
  {2\ov {{{\xi_1}^{13}}\,\xi_2}} \\&- 
  {2\ov {{{\xi_1}^9}\,\xi_2}} - {2\ov {{{\xi_1}^5}\,\xi_2}} - 
  {6\ov {\xi_1\,\xi_2}} + {{2\,{{\xi_1}^3}}\ov {\xi_2}} - 
  {{2\,{{\xi_1}^7}}\ov {\xi_2}} + 
  {{2\,\xi_2}\ov {{{\xi_1}^{23}}}} - 
  {{2\,\xi_2}\ov {{{\xi_1}^{19}}}} - 
  {{2\,\xi_2}\ov {{{\xi_1}^{15}}}} + 
  {{2\,\xi_2}\ov {{{\xi_1}^{11}}}} + 
  {{2\,\xi_2}\ov {{{\xi_1}^7}}} + 
  {{6\,\xi_2}\ov {{{\xi_1}^3}}} \\&+ 2\,\xi_1\,\xi_2 + 
  2\,{{\xi_1}^5}\,\xi_2 + 
  {{2\,{{\xi_2}^3}}\ov {{{\xi_1}^{21}}}} - 
  {{4\,{{\xi_2}^3}}\ov {{{\xi_1}^{17}}}} + 
  {{4\,{{\xi_2}^3}}\ov {{{\xi_1}^{13}}}} - 
  {{2\,{{\xi_2}^3}}\ov {{{\xi_1}^9}}} + 
  {{6\,{{\xi_2}^3}}\ov {{{\xi_1}^5}}} + 
  {{2\,{{\xi_2}^3}}\ov {\xi_1}} - 2\,{{\xi_1}^3}\,{{\xi_2}^3} + 
  {{2\,{{\xi_2}^5}}\ov {{{\xi_1}^{19}}}} - 
  {{4\,{{\xi_2}^5}}\ov {{{\xi_1}^{15}}}}\\& + 
  {{4\,{{\xi_2}^5}}\ov {{{\xi_1}^{11}}}} - 
  {{14\,{{\xi_2}^5}}\ov {{{\xi_1}^7}}} - 
  {{10\,{{\xi_2}^5}}\ov {{{\xi_1}^3}}} + 
  2\,\xi_1\,{{\xi_2}^5} + 
  {{2\,{{\xi_2}^7}}\ov {{{\xi_1}^{17}}}} - 
  {{4\,{{\xi_2}^7}}\ov {{{\xi_1}^{13}}}} + 
  {{10\,{{\xi_2}^7}}\ov {{{\xi_1}^9}}} + 
  {{10\,{{\xi_2}^7}}\ov {{{\xi_1}^5}}} - 
  {{2\,{{\xi_2}^7}}\ov {\xi_1}} + 
  {{2\,{{\xi_2}^9}}\ov {{{\xi_1}^{15}}}} - 
  {{6\,{{\xi_2}^9}}\ov {{{\xi_1}^{11}}}} \\&- 
  {{6\,{{\xi_2}^9}}\ov {{{\xi_1}^7}}} + 
  {{2\,{{\xi_2}^9}}\ov {{{\xi_1}^3}}} + 
  {{2\,{{\xi_2}^{11}}}\ov {{{\xi_1}^{13}}}} + 
  {{2\,{{\xi_2}^{11}}}\ov {{{\xi_1}^9}}}
\\ \hline
\end{array}\]
\newpage
{\bf Table 3}$\; \hfill
{_{1}{<}}  vac|\s^z|-q \xi_1,-q \xi_2>_{+-;1} =
-{_{0}{<}}vac|\s^z|-q \xi_1,-q \xi_2>_{-+;0} $ \linebreak
\hspace*{30mm}\hfill $={_{1}{<}}vac|\s^z|\xi_2^{-1},\xi_1^{-1}>_{+-;1} =
-{_{0}{<}}\xi|\s^z|\xi_2^{-1},\xi_1^{-1}>_{-+;0} $
\[\begin{array}{|l|l|}\hline
q^{\#}& {\rm coefficient} \\ \hline
1&
{{-2}\ov {{{\xi_1}^2}}} + {2\ov {{{\xi_2}^2}}}
\\ \hline
2&
2 - {2\ov {{{\xi_1}^4}}} + {2\ov {{{\xi_2}^4}}} - 
  {{2\,{{\xi_1}^2}}\ov {{{\xi_2}^2}}}
\\ \hline
3&
{{-2}\ov {{{\xi_1}^6}}} + {6\ov {{{\xi_1}^2}}} + 
  2\,{{\xi_1}^2} + {2\ov {{{\xi_2}^6}}} - {4\ov {{{\xi_2}^2}}} - 
  2\,{{\xi_2}^2} - {{2\,{{\xi_2}^2}}\ov {{{\xi_1}^4}}}
\\ \hline
4&
-4 - {2\ov {{{\xi_1}^8}}} + {4\ov {{{\xi_1}^4}}} + 
  {2\ov {{{\xi_2}^8}}} - {4\ov {{{\xi_2}^4}}} - 
  {{2\,{{\xi_1}^4}}\ov {{{\xi_2}^4}}} + 
  {2\ov {{{\xi_1}^2}\,{{\xi_2}^2}}} + 
  {{8\,{{\xi_1}^2}}\ov {{{\xi_2}^2}}} - 
  {{2\,{{\xi_2}^2}}\ov {{{\xi_1}^6}}} - 
  {{2\,{{\xi_2}^2}}\ov {{{\xi_1}^2}}}
\\ \hline
5&
{{-2}\ov {{{\xi_1}^{10}}}} + {4\ov {{{\xi_1}^6}}} - 
  {6\ov {{{\xi_1}^2}}} - 4\,{{\xi_1}^2} + 
  {2\ov {{{\xi_2}^{10}}}} - {4\ov {{{\xi_2}^6}}} + 
  {2\ov {{{\xi_1}^2}\,{{\xi_2}^4}}} + {2\ov {{{\xi_2}^2}}} + 
  6\,{{\xi_2}^2} - {{2\,{{\xi_2}^2}}\ov {{{\xi_1}^8}}} + 
  {{6\,{{\xi_2}^2}}\ov {{{\xi_1}^4}}} - 
  {{2\,{{\xi_2}^4}}\ov {{{\xi_1}^6}}} - 
  {{2\,{{\xi_2}^4}}\ov {{{\xi_1}^2}}}
\\ \hline
6&
4 - {2\ov {{{\xi_1}^{12}}}} + {4\ov {{{\xi_1}^8}}} + 
  {2\ov {{{\xi_1}^4}}} + 2\,{{\xi_1}^4} + 
  {2\ov {{{\xi_2}^{12}}}} - {4\ov {{{\xi_2}^8}}} + 
  {2\ov {{{\xi_1}^2}\,{{\xi_2}^6}}} - 
  {{2\,{{\xi_1}^6}}\ov {{{\xi_2}^6}}} + 
  {{8\,{{\xi_1}^4}}\ov {{{\xi_2}^4}}} - 
  {4\ov {{{\xi_1}^2}\,{{\xi_2}^2}}} - 
  {{6\,{{\xi_1}^2}}\ov {{{\xi_2}^2}}} \\&- 
  {{2\,{{\xi_2}^2}}\ov {{{\xi_1}^{10}}}} + 
  {{4\,{{\xi_2}^2}}\ov {{{\xi_1}^6}}} - 
  {{2\,{{\xi_2}^2}}\ov {{{\xi_1}^2}}} - 2\,{{\xi_2}^4} - 
  {{2\,{{\xi_2}^4}}\ov {{{\xi_1}^8}}} - 
  {{2\,{{\xi_2}^4}}\ov {{{\xi_1}^4}}}
\\ \hline
7&
{{-2}\ov {{{\xi_1}^{14}}}} + {4\ov {{{\xi_1}^{10}}}} + 
  {4\ov {{{\xi_1}^2}}} + 2\,{{\xi_1}^2} + 
  {2\ov {{{\xi_2}^{14}}}} - {4\ov {{{\xi_2}^{10}}}} + 
  {2\ov {{{\xi_1}^2}\,{{\xi_2}^8}}} - 
  {4\ov {{{\xi_1}^2}\,{{\xi_2}^4}}} + 
  {{2\,{{\xi_1}^2}}\ov {{{\xi_2}^4}}} + {2\ov {{{\xi_2}^2}}} + 
  {2\ov {{{\xi_1}^4}\,{{\xi_2}^2}}} \\&- 6\,{{\xi_2}^2} - 
  {{2\,{{\xi_2}^2}}\ov {{{\xi_1}^{12}}}} + 
  {{4\,{{\xi_2}^2}}\ov {{{\xi_1}^8}}} - 
  {{12\,{{\xi_2}^2}}\ov {{{\xi_1}^4}}} - 
  {{2\,{{\xi_2}^4}}\ov {{{\xi_1}^{10}}}} + 
  {{6\,{{\xi_2}^4}}\ov {{{\xi_1}^6}}} + 
  {{6\,{{\xi_2}^4}}\ov {{{\xi_1}^2}}} - 
  {{2\,{{\xi_2}^6}}\ov {{{\xi_1}^8}}} - 
  {{2\,{{\xi_2}^6}}\ov {{{\xi_1}^4}}}
\\ \hline
8&
-10 - {2\ov {{{\xi_1}^{16}}}} + {4\ov {{{\xi_1}^{12}}}} - 
  {{12}\ov {{{\xi_1}^4}}} - 4\,{{\xi_1}^4} + 
  {2\ov {{{\xi_2}^{16}}}} - {4\ov {{{\xi_2}^{12}}}} + 
  {2\ov {{{\xi_1}^2}\,{{\xi_2}^{10}}}} - 
  {{2\,{{\xi_1}^8}}\ov {{{\xi_2}^8}}} - 
  {4\ov {{{\xi_1}^2}\,{{\xi_2}^6}}} + 
  {{2\,{{\xi_1}^2}}\ov {{{\xi_2}^6}}} + 
  {{8\,{{\xi_1}^6}}\ov {{{\xi_2}^6}}} + {8\ov {{{\xi_2}^4}}}\\& + 
  {2\ov {{{\xi_1}^4}\,{{\xi_2}^4}}} - 
  {{8\,{{\xi_1}^4}}\ov {{{\xi_2}^4}}} + 
  {4\ov {{{\xi_1}^2}\,{{\xi_2}^2}}} - 
  {{2\,{{\xi_1}^2}}\ov {{{\xi_2}^2}}} - 
  {{2\,{{\xi_2}^2}}\ov {{{\xi_1}^{14}}}} + 
  {{4\,{{\xi_2}^2}}\ov {{{\xi_1}^{10}}}} - 
  {{4\,{{\xi_2}^2}}\ov {{{\xi_1}^6}}} + 
  {{14\,{{\xi_2}^2}}\ov {{{\xi_1}^2}}} + 
  2\,{{\xi_1}^2}\,{{\xi_2}^2} + 4\,{{\xi_2}^4} - 
  {{2\,{{\xi_2}^4}}\ov {{{\xi_1}^{12}}}} + 
  {{4\,{{\xi_2}^4}}\ov {{{\xi_1}^8}}}\\& + 
  {{2\,{{\xi_2}^4}}\ov {{{\xi_1}^4}}} - 
  {{2\,{{\xi_2}^6}}\ov {{{\xi_1}^{10}}}} - 
  {{2\,{{\xi_2}^6}}\ov {{{\xi_1}^6}}} - 
  {{2\,{{\xi_2}^6}}\ov {{{\xi_1}^2}}}
\\ \hline
9&
{{-2}\ov {{{\xi_1}^{18}}}} + {4\ov {{{\xi_1}^{14}}}} - 
  {6\ov {{{\xi_1}^6}}} + 2\,{{\xi_1}^2} + 2\,{{\xi_1}^6} + 
  {2\ov {{{\xi_2}^{18}}}} - {4\ov {{{\xi_2}^{14}}}} + 
  {2\ov {{{\xi_1}^2}\,{{\xi_2}^{12}}}} - 
  {4\ov {{{\xi_1}^2}\,{{\xi_2}^8}}} + 
  {{2\,{{\xi_1}^2}}\ov {{{\xi_2}^8}}} + {8\ov {{{\xi_2}^6}}} + 
  {2\ov {{{\xi_1}^4}\,{{\xi_2}^6}}} \\&+ 
  {6\ov {{{\xi_1}^2}\,{{\xi_2}^4}}} - 
  {{4\,{{\xi_1}^2}}\ov {{{\xi_2}^4}}} - {6\ov {{{\xi_2}^2}}} - 
  {6\ov {{{\xi_1}^4}\,{{\xi_2}^2}}} + 
  {{2\,{{\xi_1}^4}}\ov {{{\xi_2}^2}}} + 4\,{{\xi_2}^2} - 
  {{2\,{{\xi_2}^2}}\ov {{{\xi_1}^{16}}}} + 
  {{4\,{{\xi_2}^2}}\ov {{{\xi_1}^{12}}}} - 
  {{6\,{{\xi_2}^2}}\ov {{{\xi_1}^8}}} + 
  {{16\,{{\xi_2}^2}}\ov {{{\xi_1}^4}}} - 
  {{2\,{{\xi_2}^4}}\ov {{{\xi_1}^{14}}}} + 
  {{4\,{{\xi_2}^4}}\ov {{{\xi_1}^{10}}}}\\& - 
  {{10\,{{\xi_2}^4}}\ov {{{\xi_1}^6}}} - 
  {{12\,{{\xi_2}^4}}\ov {{{\xi_1}^2}}} - 2\,{{\xi_2}^6} - 
  {{2\,{{\xi_2}^6}}\ov {{{\xi_1}^{12}}}} + 
  {{6\,{{\xi_2}^6}}\ov {{{\xi_1}^8}}} + 
  {{6\,{{\xi_2}^6}}\ov {{{\xi_1}^4}}} - 
  {{2\,{{\xi_2}^8}}\ov {{{\xi_1}^{10}}}} - 
  {{2\,{{\xi_2}^8}}\ov {{{\xi_1}^6}}}
\\ \hline
10&
14 - {2\ov {{{\xi_1}^{20}}}} + {4\ov {{{\xi_1}^{16}}}} - 
  {8\ov {{{\xi_1}^8}}} + {{14}\ov {{{\xi_1}^4}}} + 
  {2\ov {{{\xi_2}^{20}}}} - {4\ov {{{\xi_2}^{16}}}} + 
  {2\ov {{{\xi_1}^2}\,{{\xi_2}^{14}}}} - 
  {4\ov {{{\xi_1}^2}\,{{\xi_2}^{10}}}} + 
  {{2\,{{\xi_1}^2}}\ov {{{\xi_2}^{10}}}} - 
  {{2\,{{\xi_1}^{10}}}\ov {{{\xi_2}^{10}}}} + 
  {8\ov {{{\xi_2}^8}}} + {2\ov {{{\xi_1}^4}\,{{\xi_2}^8}}} \\&+ 
  {{8\,{{\xi_1}^8}}\ov {{{\xi_2}^8}}} + 
  {6\ov {{{\xi_1}^2}\,{{\xi_2}^6}}} - 
  {{4\,{{\xi_1}^2}}\ov {{{\xi_2}^6}}} - 
  {{8\,{{\xi_1}^6}}\ov {{{\xi_2}^6}}} - 
  {{10}\ov {{{\xi_2}^4}}} - {4\ov {{{\xi_1}^4}\,{{\xi_2}^4}}} - 
  {2\ov {{{\xi_1}^2}\,{{\xi_2}^2}}} - 
  {{2\,{{\xi_2}^2}}\ov {{{\xi_1}^{18}}}} + 
  {{4\,{{\xi_2}^2}}\ov {{{\xi_1}^{14}}}} - 
  {{6\,{{\xi_2}^2}}\ov {{{\xi_1}^{10}}}} + 
  {{2\,{{\xi_2}^2}}\ov {{{\xi_1}^6}}} - 
  {{14\,{{\xi_2}^2}}\ov {{{\xi_1}^2}}} \\&- 
  4\,{{\xi_1}^2}\,{{\xi_2}^2} + 2\,{{\xi_2}^4} - 
  {{2\,{{\xi_2}^4}}\ov {{{\xi_1}^{16}}}} + 
  {{4\,{{\xi_2}^4}}\ov {{{\xi_1}^{12}}}} - 
  {{2\,{{\xi_2}^4}}\ov {{{\xi_1}^8}}} + 
  {{2\,{{\xi_2}^4}}\ov {{{\xi_1}^4}}} - 
  {{2\,{{\xi_2}^6}}\ov {{{\xi_1}^{14}}}} + 
  {{4\,{{\xi_2}^6}}\ov {{{\xi_1}^{10}}}} + 
  {{2\,{{\xi_2}^6}}\ov {{{\xi_1}^6}}} + 
  {{4\,{{\xi_2}^6}}\ov {{{\xi_1}^2}}} - 
  {{2\,{{\xi_2}^8}}\ov {{{\xi_1}^{12}}}} - 
  {{2\,{{\xi_2}^8}}\ov {{{\xi_1}^8}}} - 
  {{2\,{{\xi_2}^8}}\ov {{{\xi_1}^4}}}
\\ \hline
11&
{{-2}\ov {{{\xi_1}^{22}}}} + {4\ov {{{\xi_1}^{18}}}} - 
  {8\ov {{{\xi_1}^{10}}}} + {8\ov {{{\xi_1}^6}}} - 
  {{12}\ov {{{\xi_1}^2}}} - 6\,{{\xi_1}^2} - 4\,{{\xi_1}^6} + 
  {2\ov {{{\xi_2}^{22}}}} - {4\ov {{{\xi_2}^{18}}}} + 
  {2\ov {{{\xi_1}^2}\,{{\xi_2}^{16}}}} - 
  {4\ov {{{\xi_1}^2}\,{{\xi_2}^{12}}}} + 
  {{2\,{{\xi_1}^2}}\ov {{{\xi_2}^{12}}}}\\& + 
  {8\ov {{{\xi_2}^{10}}}} + 
  {2\ov {{{\xi_1}^4}\,{{\xi_2}^{10}}}} + 
  {6\ov {{{\xi_1}^2}\,{{\xi_2}^8}}} - 
  {{4\,{{\xi_1}^2}}\ov {{{\xi_2}^8}}} - 
  {{12}\ov {{{\xi_2}^6}}} - {4\ov {{{\xi_1}^4}\,{{\xi_2}^6}}} + 
  {2\ov {{{\xi_1}^6}\,{{\xi_2}^4}}} - 
  {6\ov {{{\xi_1}^2}\,{{\xi_2}^4}}} + 
  {{8\,{{\xi_1}^2}}\ov {{{\xi_2}^4}}} + {8\ov {{{\xi_2}^2}}} - 
  {2\ov {{{\xi_1}^8}\,{{\xi_2}^2}}} + 
  {8\ov {{{\xi_1}^4}\,{{\xi_2}^2}}} \\&- 
  {{4\,{{\xi_1}^4}}\ov {{{\xi_2}^2}}} - 
  {{2\,{{\xi_2}^2}}\ov {{{\xi_1}^{20}}}} + 
  {{4\,{{\xi_2}^2}}\ov {{{\xi_1}^{16}}}} - 
  {{6\,{{\xi_2}^2}}\ov {{{\xi_1}^{12}}}} + 
  {{8\,{{\xi_2}^2}}\ov {{{\xi_1}^8}}} - 
  {{12\,{{\xi_2}^2}}\ov {{{\xi_1}^4}}} + 
  2\,{{\xi_1}^4}\,{{\xi_2}^2} - 
  {{2\,{{\xi_2}^4}}\ov {{{\xi_1}^{18}}}} + 
  {{4\,{{\xi_2}^4}}\ov {{{\xi_1}^{14}}}} - 
  {{4\,{{\xi_2}^4}}\ov {{{\xi_1}^{10}}}} + 
  {{14\,{{\xi_2}^4}}\ov {{{\xi_1}^6}}} + 
  {{16\,{{\xi_2}^4}}\ov {{{\xi_1}^2}}} \\&+ 4\,{{\xi_2}^6} - 
  {{2\,{{\xi_2}^6}}\ov {{{\xi_1}^{16}}}} + 
  {{4\,{{\xi_2}^6}}\ov {{{\xi_1}^{12}}}} - 
  {{10\,{{\xi_2}^6}}\ov {{{\xi_1}^8}}} - 
  {{10\,{{\xi_2}^6}}\ov {{{\xi_1}^4}}} - 
  {{2\,{{\xi_2}^8}}\ov {{{\xi_1}^{14}}}} + 
  {{6\,{{\xi_2}^8}}\ov {{{\xi_1}^{10}}}} + 
  {{6\,{{\xi_2}^8}}\ov {{{\xi_1}^6}}} - 
  {{2\,{{\xi_2}^8}}\ov {{{\xi_1}^2}}} - 
  {{2\,{{\xi_2}^{10}}}\ov {{{\xi_1}^{12}}}} - 
  {{2\,{{\xi_2}^{10}}}\ov {{{\xi_1}^8}}}
\\ \hline
12&
-2 - {2\ov {{{\xi_1}^{24}}}} + {4\ov {{{\xi_1}^{20}}}} - 
  {8\ov {{{\xi_1}^{12}}}} + {{14}\ov {{{\xi_1}^8}}} - 
  {{12}\ov {{{\xi_1}^4}}} + 8\,{{\xi_1}^4} + 2\,{{\xi_1}^8} + 
  {2\ov {{{\xi_2}^{24}}}} - {4\ov {{{\xi_2}^{20}}}} + 
  {2\ov {{{\xi_1}^2}\,{{\xi_2}^{18}}}} - 
  {4\ov {{{\xi_1}^2}\,{{\xi_2}^{14}}}} + 
  {{2\,{{\xi_1}^2}}\ov {{{\xi_2}^{14}}}}\\& + 
  {8\ov {{{\xi_2}^{12}}}} + 
  {2\ov {{{\xi_1}^4}\,{{\xi_2}^{12}}}} - 
  {{2\,{{\xi_1}^{12}}}\ov {{{\xi_2}^{12}}}} + 
  {6\ov {{{\xi_1}^2}\,{{\xi_2}^{10}}}} - 
  {{4\,{{\xi_1}^2}}\ov {{{\xi_2}^{10}}}} + 
  {{8\,{{\xi_1}^{10}}}\ov {{{\xi_2}^{10}}}} - 
  {{12}\ov {{{\xi_2}^8}}} - {4\ov {{{\xi_1}^4}\,{{\xi_2}^8}}} - 
  {{8\,{{\xi_1}^8}}\ov {{{\xi_2}^8}}} + 
  {2\ov {{{\xi_1}^6}\,{{\xi_2}^6}}} - 
  {6\ov {{{\xi_1}^2}\,{{\xi_2}^6}}} + 
  {{6\,{{\xi_1}^2}}\ov {{{\xi_2}^6}}} \\&+ {8\ov {{{\xi_2}^4}}} + 
  {4\ov {{{\xi_1}^4}\,{{\xi_2}^4}}} - 
  {{6\,{{\xi_1}^4}}\ov {{{\xi_2}^4}}} - 
  {2\ov {{{\xi_1}^{10}}\,{{\xi_2}^2}}} - 
  {8\ov {{{\xi_1}^2}\,{{\xi_2}^2}}} - 
  {{4\,{{\xi_1}^2}}\ov {{{\xi_2}^2}}} + 
  {{2\,{{\xi_1}^6}}\ov {{{\xi_2}^2}}} - 
  {{2\,{{\xi_2}^2}}\ov {{{\xi_1}^{22}}}} + 
  {{4\,{{\xi_2}^2}}\ov {{{\xi_1}^{18}}}} - 
  {{6\,{{\xi_2}^2}}\ov {{{\xi_1}^{14}}}} + 
  {{6\,{{\xi_2}^2}}\ov {{{\xi_1}^{10}}}} + 
  {{6\,{{\xi_2}^2}}\ov {{{\xi_1}^6}}} \\&+ 
  {{18\,{{\xi_2}^2}}\ov {{{\xi_1}^2}}} + 
  4\,{{\xi_1}^2}\,{{\xi_2}^2} - 12\,{{\xi_2}^4} - 
  {{2\,{{\xi_2}^4}}\ov {{{\xi_1}^{20}}}} + 
  {{4\,{{\xi_2}^4}}\ov {{{\xi_1}^{16}}}} - 
  {{4\,{{\xi_2}^4}}\ov {{{\xi_1}^{12}}}} - 
  {{6\,{{\xi_2}^4}}\ov {{{\xi_1}^4}}} - 
  {{2\,{{\xi_2}^6}}\ov {{{\xi_1}^{18}}}} + 
  {{4\,{{\xi_2}^6}}\ov {{{\xi_1}^{14}}}} - 
  {{2\,{{\xi_2}^6}}\ov {{{\xi_1}^{10}}}} + 
  {{2\,{{\xi_2}^6}}\ov {{{\xi_1}^6}}} \\&- 
  {{4\,{{\xi_2}^6}}\ov {{{\xi_1}^2}}} - 2\,{{\xi_2}^8} - 
  {{2\,{{\xi_2}^8}}\ov {{{\xi_1}^{16}}}} + 
  {{4\,{{\xi_2}^8}}\ov {{{\xi_1}^{12}}}} + 
  {{2\,{{\xi_2}^8}}\ov {{{\xi_1}^8}}} + 
  {{4\,{{\xi_2}^8}}\ov {{{\xi_1}^4}}} - 
  {{2\,{{\xi_2}^{10}}}\ov {{{\xi_1}^{14}}}} - 
  {{2\,{{\xi_2}^{10}}}\ov {{{\xi_1}^{10}}}} - 
  {{2\,{{\xi_2}^{10}}}\ov {{{\xi_1}^6}}}
\\ \hline
\end{array}\]
\newpage
\section*{Figure Captions}

\noindent {\bf Fig. 1} A scatter plot of $w=E(\xi_1) + E(\xi_2)$
vs $k=-i \log (\tau(\xi_1) \tau(\xi_2))$ for $(\xi_1,\xi_2)$
randomly distributed over $0< \theta_i \leq 2\pi$ (where
$\xi_i=e^{i \theta_i}$). 
The lower limit of the points indicates the location of the
two-particle threshold.
Only the $0<k<2 \pi$ portion of the
plot is shown.

\vspace{3mm}
\noindent {\bf Fig. 2} A scatter plot of $w=E(\xi_1) + E(\xi_2)
+E(\xi_3) +E(\xi_4)$
vs $k=-i \log (\tau(\xi_1) \tau(\xi_2) \tau(\xi_3)
\tau( \xi_4))$ for $(\xi_1,\xi_2,\xi_3,\xi_4)$
randomly distributed over $0< \theta_i \leq 2\pi$ (where
$\xi_i=e^{i \theta_i}$). 
The lower limit of the points indicates the location of the
four-particle threshold.
Only the $0<k<2 \pi$ portion of the
plot is shown.

\vspace{3mm}
\noindent {\bf Fig. 3} $S_2(w,k=0)$ vs $w$ for a range of $q$ values.

\vspace{3mm}
\noindent{{\bf Fig. 4} $S_2(w,k=\pi)$ vs $w$ for a range of $q$ values.

\vspace{3mm}
\noindent {\bf Fig. 5} $S_2(w,k)$ vs $w$ for a range of 
$k$ values close to $k=0$, and for fixed $q=-0.2$.

\vspace{3mm}
\noindent {\bf Fig. 6} $S_2(w,k)$ vs $w$ for a range of 
$k$ values close to $k=\pi$, and for fixed $q=-0.2$.
\newpage
\epsfbox{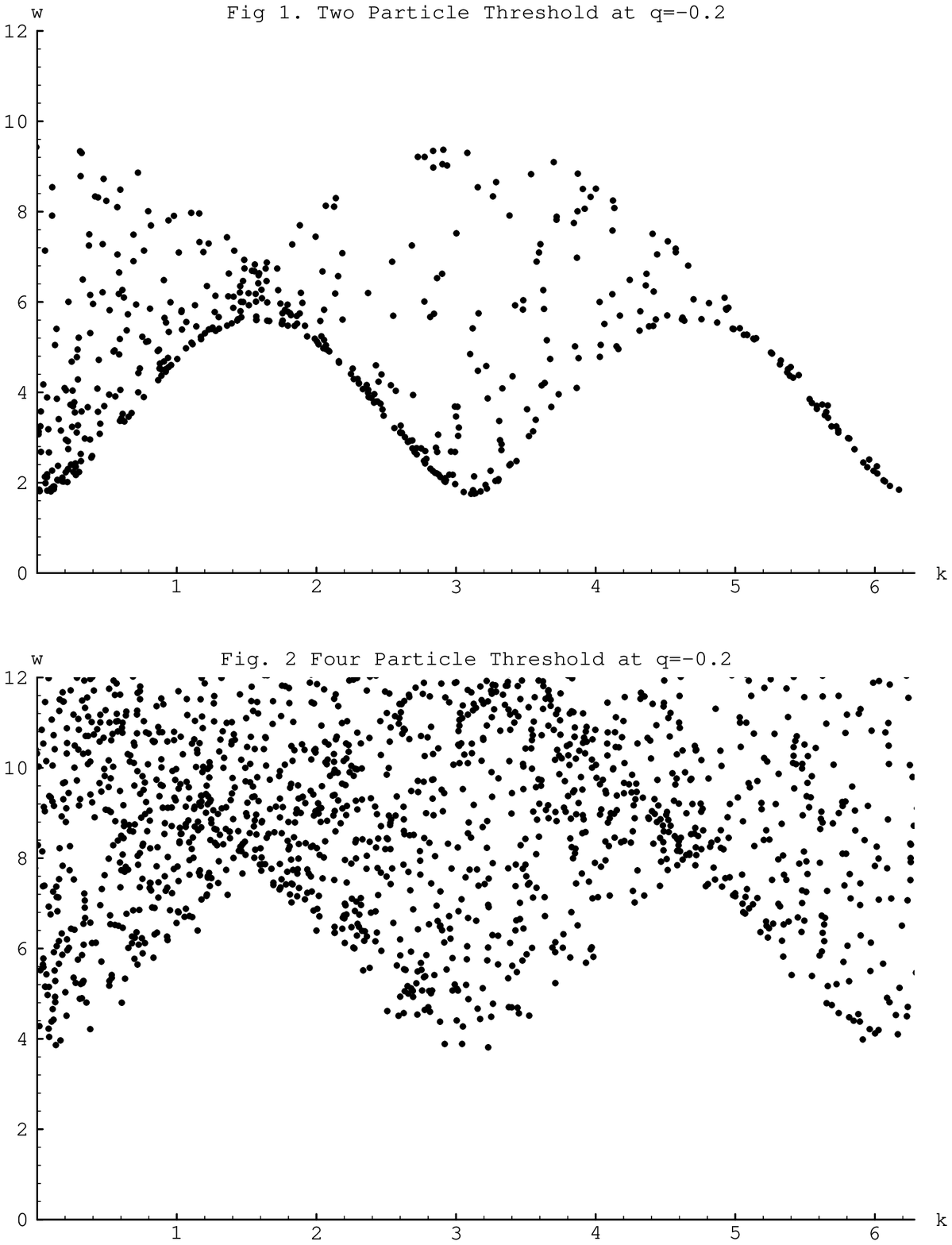}
\epsfbox{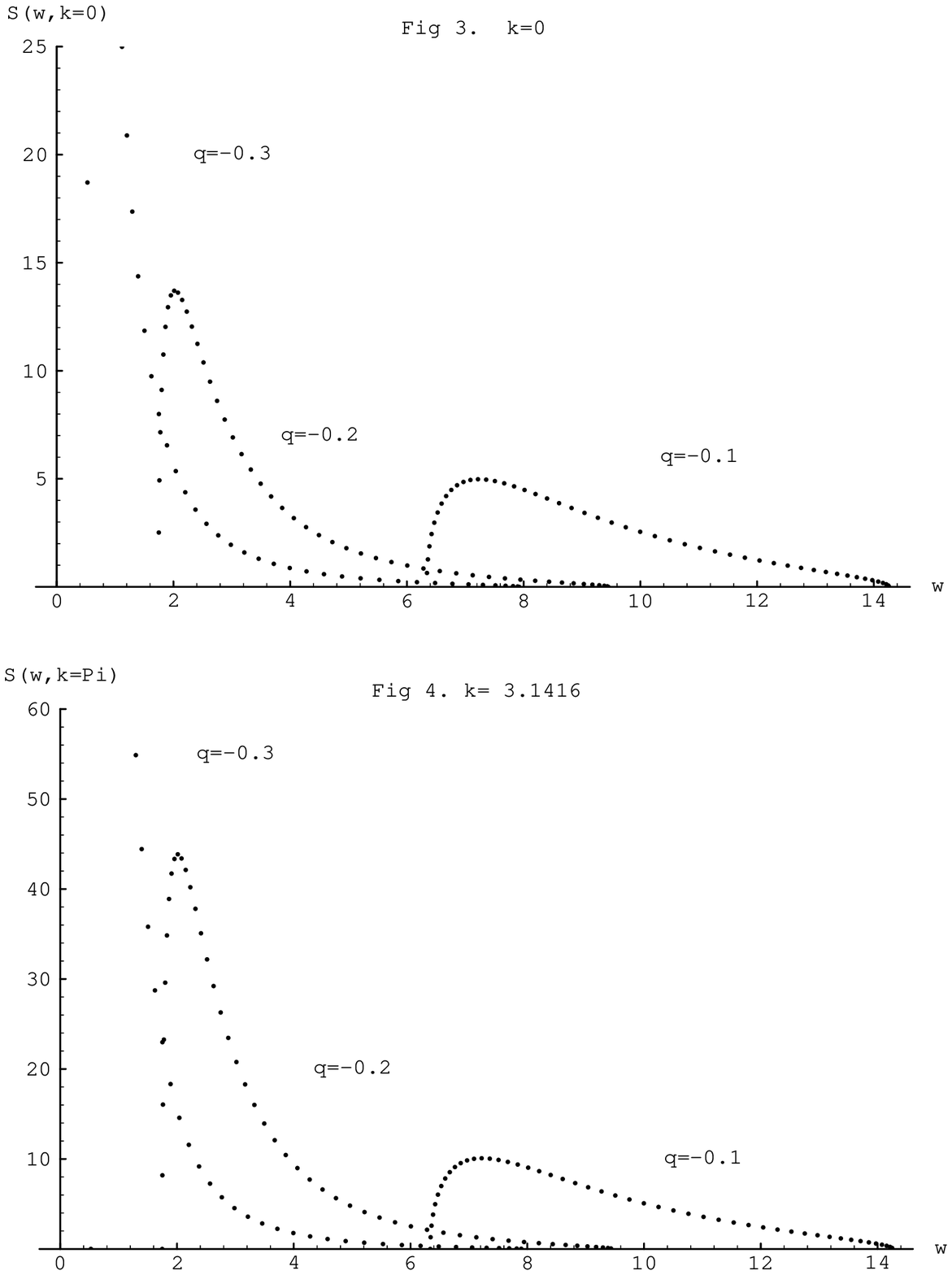}     
\epsfbox{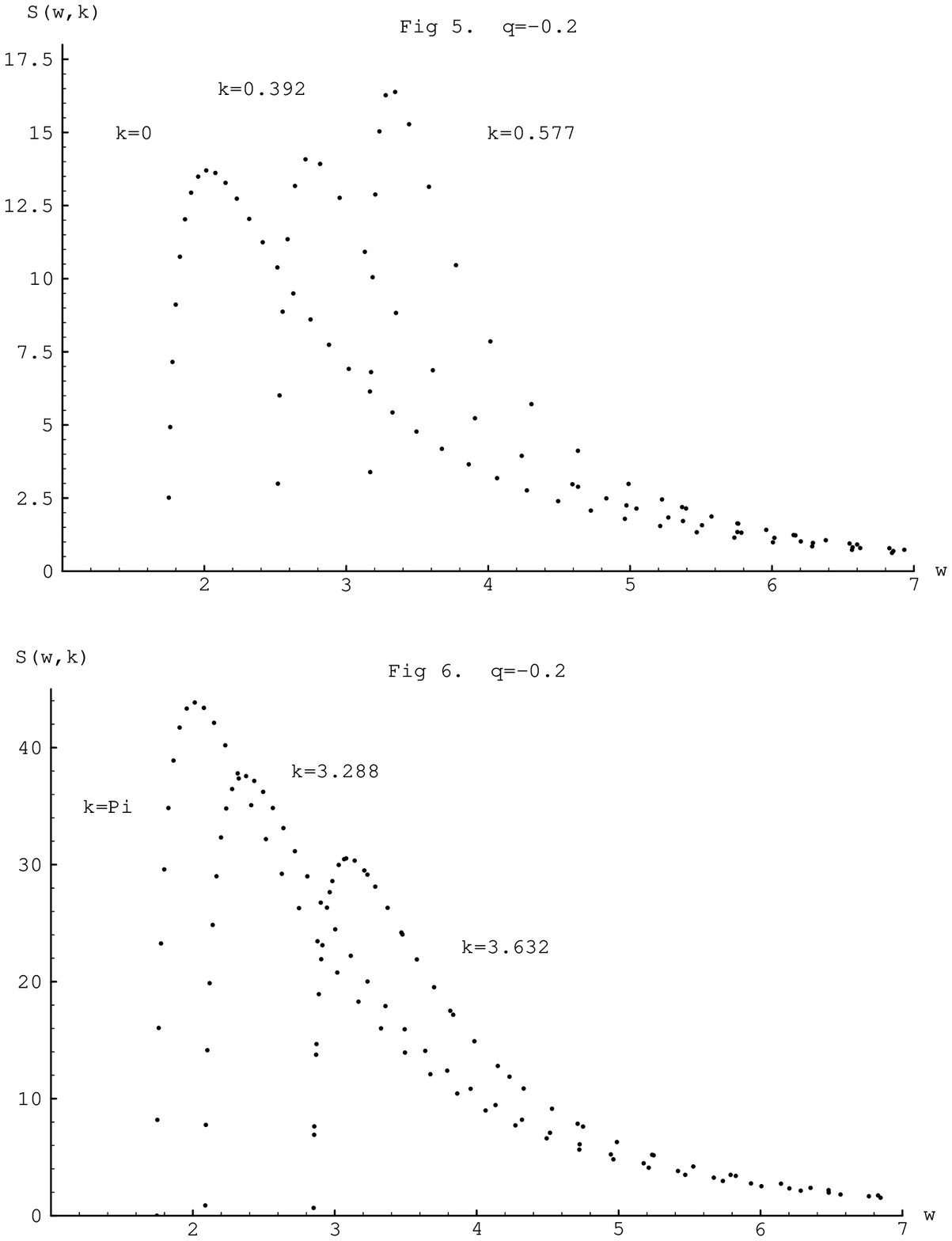}

\baselineskip=15pt
\begin{thebibliography}{10}

\bibitem{Daval92}
B.~Davies, O.~Foda, M.~Jimbo, T.~Miwa, and A.~Nakayashiki.
\newblock {\em Comm. Math. Phys.}, 151:89, 1993.

\bibitem{collin}
M.~Jimbo, K.~Miki, T.~Miwa, and A.~Nakayashiki.
\newblock Correlation functions of the {X}{X}{Z} model for {${\D}<-1$}, 1992.
\newblock RIMS preprint, hep-th/9205055.

\bibitem{idzal93}
M.~Idzumi, T.~Tokihiro, K.~Iohara, M.~Jimbo, T.~Miwa, and T.~Nakashima.
\newblock {\em Int. J. Mod. Phys.}, A8:1479, 1993.

\bibitem{Fodal93}
O.~Foda, M.~Jimbo, T.~Miwa, K.~Miki, and A.~Nakayashiki.
\newblock {\em J. Math. Phys.}, 35:13--46, 1993.

\bibitem{Jimal93b}
M.~Jimbo, T.~Miwa, and A.~Nakayashiki.
\newblock {\em J. Phys. A: Math. Gen.}, 26:2199, 1993.

\bibitem{JiMi94}
M.~Jimbo and T.~Miwa.
\newblock {\em Algebraic Analysis of Solvable Lattice Models}.
\newblock American Mathematical Society, 1994.

\bibitem{Aff89b}
I.~Affleck.
\newblock {\em J. Phys: Cond. Mat.}, 1:3047, 1989.

\bibitem{AfSo94}
E.~S. S{\o}renson and I.~Affleck.
\newblock Equal {T}ime {C}orrelations in {H}aldane {G}ap {A}ntiferromagnets,
  1994.
\newblock Preprint UBC-93-026.

\bibitem{Deial90}
J.~Deisz, M.~Jarrel, and D.L. Cox.
\newblock {\em Phys. Rev.}, B42:4869, 1990.

\bibitem{tafa79}
L.A. Takhtadzhan and L.D. Faddeev.
\newblock {\em Russ. Math. Surveys}, 34:5:11--68, 1979.

\bibitem{KiRe87}
A.~N. Kirillov and N~Yu Reshetikhin.
\newblock {\em J. Phys.}, A20:1565, 1987.

\bibitem{Aff89}
I.~Affleck.
\newblock {\em Field Theory Methods and Quantum Critical Phenomena}.
\newblock Les Houches, Session XLIX, 1988, {\it Champs, Cordes et
  Ph\'enom\`enes Critiques}, Ed. E. Br\'ezin and J. Zinn-Justin.

\bibitem{AfWe92}
I.~Affleck and R.A.Weston.
\newblock {\em Phys. Rev.}, B45:4667, 1992.

\bibitem{Babal83}
O.~Babelon, H.~de~Vega, and C.M. Viallet.
\newblock {\em Nucl. Phys.}, B220:13--34, 1983.

\bibitem{bax82}
R.~J. Baxter.
\newblock {\em Exactly Solved Models in Statistical Mechanics}.
\newblock Academic, London, 1982.

\bibitem{Bax73}
R.J. Baxter.
\newblock {\em J. Stat. Phys.}, 9:145--182, 1973.

\end{thebibliography}
\end{document}